%% file: main.tex
\documentclass[letterpaper,twocolumn,10pt]{article}
\usepackage[twoside=true, head=13pt,
     paperwidth=8.5in, paperheight=11in,
     includeheadfoot=false, columnsep=2pc,
     top=1in, bottom=1in, inner=0.75in, outer=0.75in,
     marginparwidth=2pc,heightrounded]{geometry}

\usepackage{graphicx}
\usepackage{mathptmx}
\usepackage{amssymb}
\usepackage[varqu]{zi4}
\usepackage[T1]{fontenc}
\usepackage{enumitem}
\usepackage[font={footnotesize},labelfont={footnotesize,bf},textfont={footnotesize,it}]{caption}
\usepackage{multirow}
\usepackage{xspace}
\usepackage{xcolor}
\usepackage{float}
\usepackage{subcaption}
\usepackage{booktabs}
\usepackage{url}
\usepackage{hyperref}
\usepackage{adjustbox}
\usepackage[smartEllipses]{markdown}

\usepackage{authblk}

\usepackage{enotez}

\usepackage{pgfplots}
\usepackage{pgfplotstable}
\usepgfplotslibrary{statistics}
\pgfplotsset{compat=1.18}

\definecolor{hexBlue}{HTML}{12507D}
\definecolor{hexBronze}{HTML}{B5680B}
\definecolor{hexGreen}{HTML}{388E3C}

\usepackage{tikz}
\usetikzlibrary{positioning}
\usetikzlibrary{arrows.meta,positioning,calc,shapes.geometric, arrows, backgrounds}

\newcommand{\circled}[1]{%
  \tikz[baseline=-0.8ex]{%
    \node[draw, circle, fill=none, inner sep=0.5pt, font=\scriptsize\bf] {#1};%
  }%
}

\newcommand{\ie}{\emph{i.e.,}\xspace}
\newcommand{\eg}{\emph{e.g.,}\xspace}

\def\showcomments{0}

\ifnum\showcomments=1
\newcommand{\maria}[1]{{\footnotesize\color{blue}[maria: #1]}}
\newcommand{\kostas}[1]{{\footnotesize\color{purple}[kostas: #1]}}

\newcommand{\todo}[1]{{\footnotesize\color{red}[todo: #1]}}

\else
\newcommand{\maria}[1]{}
\newcommand{\kostas}[1]{}
\newcommand{\todo}[1]{}

\fi

\newcommand{\remove}[1]{}

\newcommand{\myitem}[1]{\noindent\textbf{#1}}
\newcommand{\sys}{BGPay\xspace}

\newcommand{\prefix}{\textit{prefix}\xspace}
\newcommand{\aspath}{\textit{AS\_PATH}\xspace}
\newcommand{\bounty}{\textit{bounty}\xspace}
\newcommand{\ribCommit}{\textit{RIBCommit}\xspace}
\newcommand{\subribCommit}{\textit{subRIBCommit}\xspace}
\newcommand{\monitorReward}{\textit{monitorReward}\xspace}
\newcommand{\filterReward}{\textit{filterReward}\xspace}
\newcommand{\releaseFunction}{\textit{releaseFunction}\xspace}
\newcommand{\maskedClaim}{\textit{maskedClaim}\xspace}
\newcommand{\claim}{\textit{claim}\xspace}
\newcommand{\proofofprefixownership}{\textit{proofOfPrefixOwnership}\xspace}
\newcommand{\denylist}{\textit{denyList}\xspace}
\newcommand{\proxySet}{\textit{proxy set}\xspace}

\makeatletter
\newcommand{\printfnsymbol}[1]{%
  \textsuperscript{\@fnsymbol{#1}}%
}
\makeatother

\title{BGPay: An Incentive-Compatible Mechanism for BGP Hijack Filtering}
\author{Tomasz Sadowy~\thanks{Authors with equal contribution to this work.}} 
\author{Constantine Doumanidis\printfnsymbol{1}}
\author{Maria Apostolaki}
\affil{Princeton University}
\date{}

\begin{document}
\pagestyle{plain}
\maketitle

\input{sections/abstract}
\input{sections/intro}
\input{sections/motivation}
\input{sections/design}
\input{sections/reward_determination}
\input{sections/evaluation}

\input{sections/related_work}
\input{sections/discussion}
\input{sections/conclusion}

\bibliographystyle{plain}
\bibliography{hotnets25-template}
\clearpage
\appendix
\input{sections/appendix}

\end{document}

%% file: sections/abstract.tex
\begin{abstract}
BGP hijacking remains a persistent threat as existing defenses, including RPKI/ROV suffer from a fundamental incentive misalignment: the networks best positioned to filter malicious announcements bear operational costs but receive no direct benefit, while the victim prefix owner captures all the value.

We advocate a market-based alternative in which prefix owners post standing bounties for filtering invalid announcements of their prefixes, turning filtering from altruism into a private transaction.
Our insight is that neither a propagating hijack nor its absence can hide from public route collectors, whose committed routing tables could become an independent root of trust for releasing funds of the bounty. We build on this insight to design BGPay, an escrow protocol in which filterers and monitors commit before either reveals, and a smart contract pays out on evidence rather than on the prefix owner's judgment.

Analyzing 1K real hijack incidents, we find that today's collectors already provide enough visibility where it matters: ASes that are more important for containing the hijack are also highly visible from the public monitors. Hence, setting rewards proportionately to containment impact discourages misbehavior.

\end{abstract}

%% file: sections/intro.tex
\section{Introduction}
BGP hijacking, where an AS originates or propagates routes for IP prefixes it does not own, remains one of the Internet's most stubborn security problems. High-profile incidents continue to occur regularly, affecting major content providers~\cite{cloudflare-hijack-incident, vodafone-akamai}, financial institutions~\cite{klayswap_bgp, celerbridge_bgp}, and even government networks~\cite{hiran2013characterizing}.
The technical solutions exist. RPKI, which has seen the most deployment, allows prefix owners to cryptographically sign Route Origin Authorizations (ROAs), and Route Origin Validation (ROV) enables routers to reject announcements inconsistent with these ROAs. Yet deployment remains sluggish: as of 2026, only roughly 24\% of ASes perform ROV~\cite{li-2023-rov}, and coverage is uneven across regions and tiers.
The root cause is economic and incentives-related, not technical. An AS that deploys ROV incurs operational costs (configuration, monitoring, risk of misconfiguration dropping legitimate traffic) but primarily protects other networks' prefixes.
Worse, filtering means giving up revenue. So the incentive isn't neutral; it's actively misaligned. This is a textbook externality problem, and it explains why voluntary adoption has been slow despite years of advocacy.

We propose abandoning the appeal to collective good and paying for the action directly. A prefix owner posts a standing bounty: ``any AS that suppresses a hijack of my prefix gets paid proportionately to how critical their filtering is to the containment of the hijack''. Filtering becomes a service with a buyer and a seller rather than a civic duty. This transforms BGP security from an act of altruism into a private transaction, aligning the incentives of filtering ASes with those of prefix owners. 
While intuitive, realizing this idea is challenging. ASes do not trust each other, the filtering AS needs assurance that if it filters, it will get paid; the prefix owner needs assurance that funds are only released for genuine filtering but the act of filtering itself destroys the evidence. Worse yet, such a mechanism could easily create the opposite effect by motivating more BGP hijacks to increase the market for filtering. Three insights make our proposed protocol, \sys, viable despite these challenges.

\myitem{Proof of filtering by absence.} Our first insight is that filtering leaves a shadow even though it leaves no trace. BGP is a flawed protocol but a generous one: it emits a continuous, redundant signal about how every AS behaves, seen from many angles at once. Where an advertisement travels is determined by business relationships fixed long before any attack, and those relationships are legible in the routing tables an AS produces every day. So the paths a hijack takes, and the paths it conspicuously fails to take, are both predictable from an AS's ordinary behavior. Hundreds of ASes already publish exactly this record and have done so for years, altruistically, to RouteViews~\cite{routeviews}, RIPE RIS~\cite{ripe-ris}. This gives the mechanism a natural asymmetry. An AS that keeps forwarding a hijack while claiming a bounty has to hope that none of those collectors saw it, when its own history of exports says several should have. Manufacturing a clean record instead would mean withholding routes from collectors long enough in advance to look ordinary, and would cost more effort than simply filtering the hijack in the first place. The cheapest way to appear to have filtered is to filter.

\myitem{Attestability and impact are aligned.} Our second insight is that absence is louder when it matters. An AS is widely observed for the same reason it can help mitigate a hijack effectively: both follow from exporting to a large customer cone. As an intuition, a provider of the hijacker is the single most valuable filterer and also the one that can least plausibly propagate an advertisement and not be seen by the various BGP monitors. A stub customer, on the other end, can lie freely but would also not be necessary for mitigating the hijack. By setting payments according to impact, we essentially leverage the Internet's hierarchy as a policing mechanism instead of relying on a central authority deciding who is trustworthy. Indeed, we find that 
the provider of a hijacker is roughly three orders of magnitude more important in constraining the hijack compared to her customer and almost two orders of magnitude more likely to get contradicted by at least one monitor if it tries to cheat.

\myitem{Blockchains provide the flexibility and accountability needed.}
Our third insight is that a mechanism written down in public needs no trust to run and no authority to govern. Escrow settles the two-sided assurance problem: funds are locked before the hijack and released by a rule grounded on the BGP-monitor's view rather than by the victim's goodwill, so neither party has to believe the other. Commitments settle the copycat problem: a filterer publishes evidence that is sealed but binding, so an AS that hears of a hijack secondhand has nothing to copy, and nobody can deny or quietly edit what they committed to once claims open. Because a bounty is a contract its author writes, each prefix owner sets its own price, its own evidence threshold, and its own trusted monitors. Owners can be as demanding as they like; ASes answer by choosing which prefixes are worth their filters. That is a market, not a standard. Finally, every claim and payout is permanent and public, which answers the worry that we might simply grow the demand for hijacks: a coalition can stage an attack and collect on it once, but an AS that keeps appearing in incidents nobody else corroborates is visible to every bounty setter thereafter. Staging a hijack has to work every time. Being caught has to work once.

The central question, then, is whether today's monitoring infrastructure provides enough visibility for this mechanism to work. In this paper, we outline the necessary conditions for filtering ASes to claim bounties and evaluate them via AS-level simulations across a dataset of real hijack incidents.
We find that provider ASes to hijackers - the exact ASes where filtering is most critical to contain the hijack - in the mean case have 648 monitors that can attest to their filtering.
Conversely, should the same ASes attempt to fraudulently claim a bounty without actually filtering a longer-prefix hijack, at least one monitor can witness their misbehavior in 97\% of cases.  Our findings suggest that \sys can be effectively deployed with today's route monitoring infrastructure. 

%% file: sections/motivation.tex
\section{Motivation}
\label{sec:motivation}
\input{figs/motivating_example_v4}

In this section, we discuss how misaligned incentives affect the deployment of filtering, and how our protocol addresses them, through the toy example in Figure~\ref{fig:motivating-example}. 

\subsection{The problem with filtering}
 Consider the 10 ASes with the peering relationships shown. AS H performs a subprefix hijack against a prefix owned by AS V, announcing a more specific prefix \texttt{1.0.0.0/24}. If all neighboring ASes follow default BGP, they route traffic for \texttt{1.0.0.0/24} to the attacker AS H. The attack fails only if AS F1, AS F2, and AS F3 detect that AS H is lying and refuse to propagate the announcement. Yet today, filtering relies on the goodwill of ASes, even as their own incentives, cost, risk, and forgone revenue, pull them toward doing nothing. First, catching the hijack requires up-to-date filters: each AS must track which prefixes its peers, customers, and providers may legitimately advertise, or at least stay in sync with recent ROAs. Second, a mistake is costly: if an ROA is wrong or a mapping is stale, which is very common~\cite{smart-rpki,hlavacek2020disco}, filtering a legitimate announcement can blackhole real traffic. Third, filtering can mean forgoing revenue: AS F1, for instance, is paid by AS H for the traffic it carries, so dropping their announcement actually costs it money. Even if AS F1, AS F2, and AS F3 filter despite all this, the benefit accrues to AS V, an AS with which they have no relationship. The result is a textbook externality: those who bear the cost are not those who reap the reward, and so many ASes simply do not filter, as the continued effectiveness of hijacks attests. 
Critically, containment is fragile: if even one of AS F1, AS F2, or AS F3 declines to filter, the hijacked route still propagates through that AS, letting the attacker draw off at least some of the traffic. For instance, if AS F1 did not filter, AS C's traffic for \texttt{1.0.0.0/24} would be redirected to AS H.

\subsection{\sys an incentives-aligned mechanism}
The example outlined above motivates a mechanism that rewards ASes for filtering, giving those that would not otherwise act a concrete reason to do so.
Next, we explain how our protocol, \sys could have provided the missing incentives, effectively mitigating the hijack.
Suppose victim AS V participates in \sys, hence posts a bounty (Fig.~\ref{fig:motivating-example}: \circled{2}) to a public blockchain ledger (\eg Ethereum), committing to pay \$x to any AS that helps prevent a BGP hijack of its prefix. Suppose that AS F1 notices the standing bounty, and, in anticipation of the rewards, installs a filter for that prefix. When the hijack occurs, AS F1 detects and filters it (\circled{3}), and immediately posts a claim to the chain, which should result in a direct payment. 
Instead of relying on mutual trust or assuming that ASes would naturally behave according to the protocol, \sys provides the necessary guarantees: \emph{(i)} AS V needs some guarantee that AS F1 really filtered; and \emph{(ii)}  AS F1 needs some guarantee that AS V will actually pay. To achieve this, \sys leverages public BGP monitors \eg RouteViews~\cite{routeviews} and RIPE RIS~\cite{ripe-ris} peers, which cover a significant portion of the Internet and can independently attest as to what really happened.
Observe, for instance, the monitor AS M1 in our example: had AS F1 not filtered, the hijacked route would have propagated onward and AS M1 would have heard \texttt{(1.0.0.0/24, [F1, H])}.
By having BGP monitors commit their RIBs to the chain, \sys allows everyone to verify by absence that AS F1 must have filtered, but also makes the release of the funds subject to that check rather than AS V's discretion. In effect, AS V does not need to trust AS F1's word and vice versa. Proof by absence is weaker than a positive proof of action, but three properties make it sufficient here.
First, \sys scales the reward by detectability, as we describe in \S\ref{sec:reward_determination}. The ASes whose filtering matters most are the ASes whose propagation is most widely observed, hence verifiable, so lying is hardest exactly where the payout is largest. 
Second, claims are bound before they are public: filterers commit to a hash of the announcement they observed (\circled{4}) and monitors commit to their RIBs (\circled{5}), both before either is revealed (\circled{6}, \circled{7}). For instance, AS F1 commits an irreversible hash of the claim to the blockchain, another AS, say AS F4, would not be able to simply copy a legitimate claim. While they can still learn about the hijack out-of-band and submit a false claim, they will be slower.
Third, residual risk is priced rather than eliminated: the bounty setter specifies the exclusion list, and the reward weighting, and can even define custom payment policies, therefore choosing its own point on the false-positive/false-negative curve.

To reduce the load on the blockchain, ease the pressure on monitors to process data fast, and remove the privacy implications of publishing full RIBs, \sys has monitors only commit a cryptographically protected hash of their RIBs (\circled{5}) and, after some time, a relevant subset of advertisements (\circled{7}). Specifically, monitors report advertisements whose \texttt{AS\_PATH} contains the claimant AS and its neighbor that forwarded the hijack to it, or advertisements where \texttt{AS\_PATH} contains the claimant AS and \texttt{prefix} matches the hijacked prefix. Because the initial commit is cryptographically protected \eg with a merkle root, the monitors cannot change their view after they commit.

While it is possible that an AS, say AS F4, claims a hijack that never happens or even colludes with an AS to hijack such that they can claim the bounty, their actions will be visible on the blockchain, hence they will be unable to do it repeatedly. An AS that repeatedly claims hijacks that no one else noticed, or claimed, can be excluded from bounties: this can be included in the conditions of the bounty. ASes thus have no real incentive to repeatedly hijack.

%% file: figs/motivating_example_v4.tex
\tikzstyle{nodeStyle} = [
    circle,
    draw,
    fill=blue!20,
    minimum size=2.5em,
    text centered,
    font=\sffamily\scriptsize,
    inner sep=0.25em
]
\tikzstyle{edgeStyle} = [->]
\tikzstyle{edgeStyleN} = [-]

\tikzstyle{sectionTitle} = [
    font=\sffamily\bfseries
]
\tikzstyle{subsectionTitle} = [
    font=\sffamily\bfseries\small
]
\tikzstyle{blockStyle} = [
    draw,
    rounded corners=3pt,
    thick,
    fill=gray!8,
    minimum width=41em,
    minimum height=4em,
    anchor=north west
]
\tikzstyle{txStyle} = [
    draw,
    rounded corners=4pt,
    thick,
    fill=orange!20,
    minimum width=3.6em,
    minimum height=3.6em,
    anchor=north west
]
\tikzstyle{blockHeader} = [
    font=\sffamily\bfseries\small
]
\tikzstyle{chainMsgStyle} = [
    draw,
    rounded corners=1.5pt,
    fill=white,
    inner sep=1pt,
    font=\sffamily\scriptsize,
    align=left,
    anchor=north west,
    minimum height=1.35em,
    draw=none,
]
\tikzstyle{bgpMsgStyle} = [
    draw,
    rounded corners=2pt,
    fill=white,
    inner sep=3pt,
    font=\sffamily\footnotesize,
    align=left,
    anchor=north west,
    minimum height=1.35em,
    dashed
]
\tikzstyle{oobMsgStyle} = [
    draw,
    rounded corners=2pt,
    fill=white,
    inner sep=3pt,
    font=\sffamily\footnotesize,
    align=left,
    anchor=north west,
    minimum height=1.35em,
    dotted
]
\tikzstyle{ribBlock} = [
    rounded corners=2pt,
    fill=gray!10,
    inner sep=3pt,
    font=\sffamily\scriptsize,
    align=left,
    anchor=north west,
    minimum height=1.35em,
    minimum width=8em,
]
\tikzstyle{timeArrow} = [
    very thick,
    -Latex
]
\tikzstyle{periodBrace} = [
    thick
]
\begin{figure*}[ht!]
    \centering
    \resizebox{0.85\linewidth}{!}{
    \begin{tikzpicture}[node distance=4em and 5em, on grid]


        \node[sectionTitle] (blockTitle) {Blockchain};
        \node[blockStyle, right=6em of blockTitle, yshift=0em, xshift=1em, anchor=west] (blockchain) {};

        \draw[timeArrow]
            ([xshift=0em,yshift=1em]blockchain.north west)
            --
            ([xshift=40em,yshift=1em]blockchain.north west)
            node[midway, left=0em, above=0.25em, font=\sffamily\bfseries] {Time};
            

        \node[txStyle, right=0.5em of blockchain.north west, yshift=-0.2em, xshift=0em, anchor=north west] (bounty) {bounty};

        \node[txStyle, fill=blue!20, right=0.2em of bounty.north east, yshift=0em, xshift=0em, anchor=north west] (maskedClaims1) {maskedClaims};
        \node[txStyle, fill=blue!20, right=0.2em of maskedClaims1.north east, yshift=0.2em, xshift=0em, anchor=north east] (maskedClaims2) {maskedClaims};
        \node[txStyle, fill=blue!20, right=0.2em of maskedClaims2.north east,yshift=0.2em, xshift=0em, anchor=north east] (maskedClaims3) {maskedClaims};

        \node[txStyle, fill=green!20, right=0.5em of maskedClaims1.north east, yshift=0em, xshift=0em, anchor=north west] (ribCommits1) {RIBCommits};
        \node[txStyle, fill=green!20, right=0.2em of ribCommits1.north east, yshift=0.2em, xshift=0em, anchor=north east] (ribCommits2) {RIBCommits};
        \node[txStyle, fill=green!20, right=0.2em of ribCommits2.north east,yshift=0.2em, xshift=0em, anchor=north east] (ribCommits3) {RIBCommits};

        \node[txStyle, fill=blue!20, right=0.5em of ribCommits1.north east, yshift=0em, xshift=0em, anchor=north west] (claims1) {claims};
        \node[txStyle, fill=blue!20, right=0.2em of claims1.north east, yshift=0.2em, xshift=0em, anchor=north east] (claims2) {claims};
        \node[txStyle, fill=blue!20, right=0.2em of claims2.north east,yshift=0.2em, xshift=0em, anchor=north east] (claims3) {claims};

        \node[txStyle, fill=green!20, right=0.5em of claims1.north east, yshift=0em, xshift=0em, anchor=north west] (subRIBCommits1) {subRIBCommits};
        \node[txStyle, fill=green!20, right=0.2em of subRIBCommits1.north east, yshift=0.2em, xshift=0em, anchor=north east] (subRIBCommits2) {subRIBCommits};
        \node[txStyle, fill=green!20, right=0.2em of subRIBCommits2.north east,yshift=0.2em, xshift=0em, anchor=north east] (subRIBCommits3) {subRIBCommits};

        \node[txStyle, right=0.5em of subRIBCommits1.north east, yshift=0em, xshift=0em, anchor=north west] (releaseFunc) {bounty.releaseFunction()};


        \node[sectionTitle, below=6em of blockTitle.south west, anchor=north west] (topoTitle) {AS Topology};
        
        \node[nodeStyle, fill=orange!20, right=3em of topoTitle.south east, yshift=2em, anchor=north west] (v) {AS V};
        \node[nodeStyle, fill=orange!20, below=5.5em of v, xshift=-0.5em] (C) {AS C};

        \node[nodeStyle, right=6em of v, yshift=-2em] (F1) {AS F1};
        \node[nodeStyle, fill=green!20, above=4em of F1, xshift=-3em] (m1) {AS M1};
        \node[nodeStyle, fill=red!20, below=5em of F1, xshift=0em] (hijacker) {AS H};

        \node[nodeStyle, right=4em of F1, yshift=-0.5em] (F3) {AS F3};
        
        \node[nodeStyle, right=0.5em of F3, yshift=4em] (F4) {AS F4};
        \node[nodeStyle, right=1em of F3, yshift=-4em] (F2) {AS F2};
        \node[nodeStyle, fill=green!20, above=3em of F2, xshift=4em] (m2) {AS M2};
        \node[nodeStyle, fill=green!20, right=5em of F4] (m3) {AS M3};

        \draw[edgeStyleN] (C) -- (v);
        \draw[edgeStyleN] (F1) -- (m1);
        \draw[edgeStyle] (F1) -- (C);
        \draw[edgeStyle] (F1) -- (hijacker);
        \draw[edgeStyleN] (hijacker) -- (F2);
        \draw[edgeStyle] (hijacker) -- (F3);
        \draw[edgeStyle] (F2) -- (m2);
        \draw[edgeStyle] (F3) -- (m2);
        \draw[edgeStyle] (F3) -- (m3);
        \draw[edgeStyle] (F3) -- (F4);
        \draw[edgeStyleN] (m1) -- (F4);

        \draw[->, thick, draw=blue!60, ] (v) to[bend left=-20] node[chainMsgStyle, fill=blue!10, pos=0.5, anchor=center, xshift=-0.5em] (legitAnn) {V/8:V} (C);

        \draw[->, thick, draw=orange!90, ] (v) to[bend left=20] node[draw=black, circle, fill=white, inner sep=0.5pt,font=\Large\bf, xshift=-1em, yshift=-0.5em] {2} (bounty);

        \draw[->, thick, draw=blue!60, ] (C) to[bend left=20] node[chainMsgStyle, fill=blue!10, pos=0.5, anchor=center, xshift=0em, yshift=-0.5em] {V/8:C,V} (F1);

        \draw[->, thick, draw=blue!60, ] (F1) to[bend left=-20] node[chainMsgStyle, fill=blue!10, pos=0.5, anchor=center, xshift=-1em, yshift=-0.5em] {V/8:F1,C,V} (m1);

        \draw[->, thick, draw=blue!60, ] (F1) to[bend left=-20] node[chainMsgStyle, fill=blue!10, pos=0.5, anchor=center, xshift=0em, yshift=0em] {V/8:F1,C,V} (hijacker);

        \draw[->, thick, draw=blue!60, ] (hijacker) to[bend left=-20] node[chainMsgStyle, fill=blue!10, pos=0.5, anchor=center, xshift=0em, yshift=-1em] {V/8:H,F1,C,V} (F2);
        \draw[->, thick, draw=blue!60, ] (hijacker) to[bend left=-20] (F3);

        \draw[->, thick, draw=blue!60, ] (F2) to[bend left=-20] node[chainMsgStyle, fill=blue!10, pos=0.5, anchor=center, xshift=1.5em, yshift=-0.5em] {V/8:F2,H,F1,C,V} (m2);

        \draw[->, thick, draw=blue!60, ] (F3) to[bend left=20] node[chainMsgStyle, fill=blue!10, pos=0.5, anchor=center, xshift=2em, yshift=-1.5em] {V/8:F3,H,F1,C,V} (m3);
        \draw[->, thick, draw=blue!60, ] (F3) to[bend left=20] (m2);
        \draw[->, thick, draw=blue!60, ] (F3) to[bend left=20] (F4);

        \node[draw,circle, fill=none, inner sep=0.5pt,font=\Large\bf] at ([xshift=-0.75em,yshift=-0.6em]legitAnn.north west) {1};

        \node[draw,circle, fill=none, inner sep=0.5pt,font=\Large\bf] at ([xshift=-0.75em,yshift=-0.6em]legitAnn.north west) {1};

        \node[draw=none, fill=none] at ([xshift=2em,yshift=-3em]hijacker.south) {Before Hijack};

        
        \node[nodeStyle, fill=orange!20, right=27em of topoTitle.south east, yshift=2em, anchor=north west] (v) {AS V};
        \node[nodeStyle, fill=orange!20, below=5.5em of v, xshift=-0.5em] (C) {AS C};

        \node[nodeStyle, right=6em of v, yshift=-2em] (F1) {AS F1};
        \node[nodeStyle, fill=green!20, above=4em of F1, xshift=-3em] (m1) {AS M1};
        \node[nodeStyle, fill=red!20, below=5em of F1, xshift=0em] (hijacker) {AS H};

        \node[nodeStyle, right=4em of F1, yshift=-0.5em] (F3) {AS F3};
        
        \node[nodeStyle, right=0.5em of F3, yshift=4em] (F4) {AS F4};
        \node[nodeStyle, right=1em of F3, yshift=-4em] (F2) {AS F2};
        \node[nodeStyle, fill=green!20, above=3em of F2, xshift=4em] (m2) {AS M2};
        \node[nodeStyle, fill=green!20, right=5em of F4] (m3) {AS M3};

        \draw[edgeStyleN] (C) -- (v);
        \draw[edgeStyleN] (F1) -- (m1);
        \draw[edgeStyle] (F1) -- (C);
        \draw[edgeStyle] (F1) -- (hijacker);
        \draw[edgeStyleN] (hijacker) -- (F2);
        \draw[edgeStyle] (hijacker) -- (F3);
        \draw[edgeStyle] (F2) -- (m2);
        \draw[edgeStyle] (F3) -- (m2);
        \draw[edgeStyle] (F3) -- (m3);
        \draw[edgeStyle] (F3) -- (F4);
        \draw[edgeStyleN] (m1) -- (F4);


        \draw[->, thick, draw=red!90] (hijacker) to[bend left=20] node[chainMsgStyle, fill=red!10, pos=0.5, anchor=center, xshift=0em, yshift=0em] (hijack) {V/24:H} (F1);
        \draw[->, thick, draw=red!90, ] (hijacker) to[bend left=-20] node[chainMsgStyle, fill=red!10, pos=0.5, anchor=center, xshift=0em, yshift=-1em] {V/24:H} (F2);
        \draw[->, thick, draw=red!90, ] (hijacker) to[bend left=-20] node[chainMsgStyle, fill=red!10, pos=0.5, anchor=center, xshift=0em, yshift=0em] {V/24:H} (F3);

        \draw[->, thick, draw=orange!90] (F1) to[bend left=-5] node[draw=black, circle, fill=white, inner sep=0.5pt,font=\Large\bf, xshift=0em, yshift=0em, pos=1.5] {6} ([xshift=-1em, yshift=0em]F1.west);
        \draw[->, thick, draw=orange!90] (F1) to[bend left=-5] node[draw=black, circle, fill=white, inner sep=0.5pt,font=\Large\bf, xshift=0em, yshift=0em, pos=1.5] {4} ([xshift=-1em, yshift=1em]F1.west);

        \draw[->, thick, draw=orange!90] (F2) to[bend left=-5] node[draw=black, circle, fill=white, inner sep=0.5pt,font=\Large\bf, xshift=0em, yshift=0em, pos=1.5] {4} ([xshift=1em, yshift=-0em]F2.east);
        \draw[->, thick, draw=orange!90] (F2) to[bend left=-5] node[draw=black, circle, fill=white, inner sep=0.5pt,font=\Large\bf, xshift=0em, yshift=0em, pos=1.5] {6} ([xshift=1em, yshift=-1em]F2.east);

        \draw[->, thick, draw=orange!90] (F3) to[bend left=-5] node[draw=black, circle, fill=white, inner sep=0.5pt,font=\Large\bf, xshift=0em, yshift=0em, pos=1.5] {6} ([xshift=-0.5em, yshift=1.25em]F3.west);
        \draw[->, thick, draw=orange!90] (F3) to[bend left=-5] node[draw=black, circle, fill=white, inner sep=0.5pt,font=\Large\bf, xshift=0em, yshift=0em, pos=1.5] {4} ([xshift=-0em, yshift=2.25em]F3.west);

        \draw[->, thick, draw=orange!90] (m1) to[bend left=-5] node[draw=black, circle, fill=white, inner sep=0.5pt,font=\Large\bf, xshift=0em, yshift=0em, pos=1.5] {7} ([xshift=-1em, yshift=0em]m1.west);
        \draw[->, thick, draw=orange!90] (m1) to[bend left=-5] node[draw=black, circle, fill=white, inner sep=0.5pt,font=\Large\bf, xshift=0em, yshift=0em, pos=1.5] {5} ([xshift=-1em, yshift=1em]m1.west);

        \draw[->, thick, draw=orange!90] (m2) to[bend left=-5] node[draw=black, circle, fill=white, inner sep=0.5pt,font=\Large\bf, xshift=0em, yshift=0em, pos=1.5] {5} ([xshift=1em, yshift=-0em]m2.east);
        \draw[->, thick, draw=orange!90] (m2) to[bend left=-5] node[draw=black, circle, fill=white, inner sep=0.5pt,font=\Large\bf, xshift=0em, yshift=0em, pos=1.5] {7} ([xshift=1em, yshift=-1em]m2.east);

        \draw[->, thick, draw=orange!90] (m3) to[bend left=-5] node[draw=black, circle, fill=white, inner sep=0.5pt,font=\Large\bf, xshift=0em, yshift=0em, pos=1.5] {5} ([xshift=1em, yshift=-0em]m3.east);
        \draw[->, thick, draw=orange!90] (m3) to[bend left=-5] node[draw=black, circle, fill=white, inner sep=0.5pt,font=\Large\bf, xshift=0em, yshift=0em, pos=1.5] {7} ([xshift=1em, yshift=-1em]m3.east);

        \node[draw,circle, fill=white, inner sep=0.5pt,font=\Large\bf] at ([xshift=-0.75em,yshift=-0.6em]hijack.north west) {3};

        \node[draw,circle, fill=white, inner sep=0.5pt,font=\Large\bf] at ([xshift=0em,yshift=0em]maskedClaims1.south) {4};
        \node[draw,circle, fill=white, inner sep=0.5pt,font=\Large\bf] at ([xshift=0em,yshift=0em]ribCommits1.south) {5};
        \node[draw,circle, fill=white, inner sep=0.5pt,font=\Large\bf] at ([xshift=0em,yshift=0em]claims1.south) {6};
        \node[draw,circle, fill=white, inner sep=0.5pt,font=\Large\bf] at ([xshift=0em,yshift=0em]subRIBCommits1.south) {7};
        \node[draw,circle, fill=white, inner sep=0.5pt,font=\Large\bf] at ([xshift=0em,yshift=0em]releaseFunc.south) {8};


        \node[draw=none, fill=none] at ([xshift=2em,yshift=-3em]hijacker.south) {After Hijack};

    \end{tikzpicture}
    }
    \caption{The graph on the left shows a miniature AS-level topology before AS H performs a hijack against a prefix of AS V. During that time, \protect\circled{1} AS V advertises its prefix (V/8) and \protect\circled{2} posts a bounty to the blockchain committing to pay some amount to ASes that help mitigate possible hijacks. We assume this triggered AS F1, AS F2, and AS F3 (among other ASes) to install appropriate filters. The graph on the right shows the topology after \protect\circled{3} AS H performs the hijack, which AS F1, AS F2, and AS F3 filter. Next, \protect\circled{4} AS F1, AS F2, and AS F3 submit maskedClaims to the blockchain \ie requests on V's bounty which do not expose enough information to allow an imposter AS to also request the payment without having observed the hijack. In response to the maskedClaims, monitors AS M1, AS M2, and AS M3 submit ribCommits, \ie a hashed and signed version of their view of BGP advertisements without knowing what exactly happened. Next, \protect\circled{6} AS F1, AS F2, and AS F3 reveal their claims, and  \protect\circled{7} the monitors AS M1, AS M2, and AS M3 publish subRIBCommits \ie a plain (not just hash) subset of their view related to the affected prefix. Finally, \protect\circled{8} the releaseFunction distributes bounty rewards.}
    \label{fig:motivating-example}
\end{figure*}
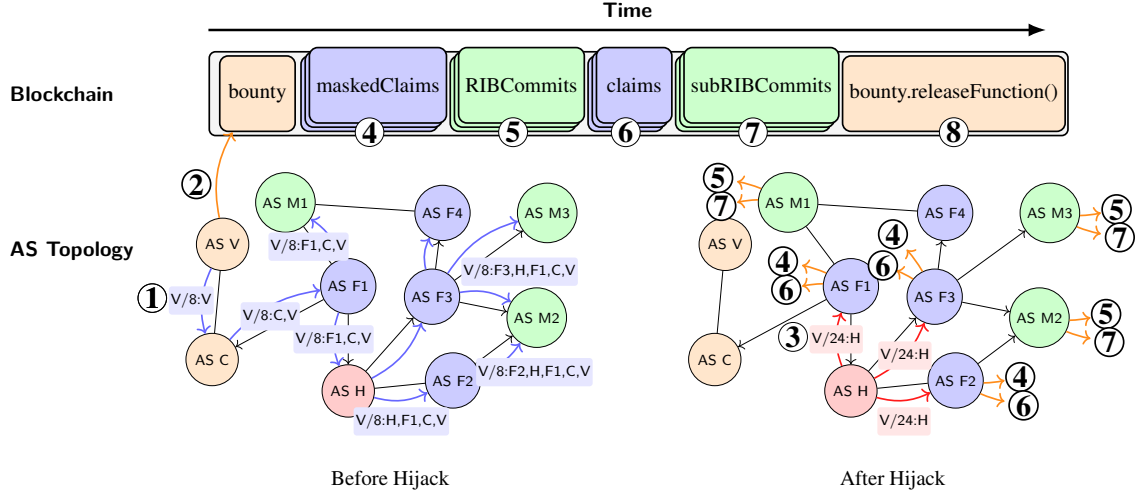

%% file: sections/design.tex
\section{Protocol Design}
\label{sec:design}
We now present the \sys protocol in detail. We first describe in \S~\ref{sec:setup} the entities participating in \sys, and the assumptions under which \sys operates. In \S~\ref{sec:description}, we walk through the protocol execution phases from bounty publication up to reward distribution and provide details on the messages \sys participants use in their protocol interactions. Finally, \S~\ref{sec:properties} discusses the properties provided by \sys and the design mechanisms that achieve them.

\subsection{Setting}

\label{sec:setup}
\myitem{Participants:} \sys participants are Autonomous Systems that take on the roles of bounty setters, filtering ASes, and monitors. Bounty setters post bounties that reward filtering invalid BGP announcements of prefixes they manage. Filtering ASes filter invalid BGP announcements they observe, and claim bounty rewards. Finally, monitors publish parts of their BGP routing tables as evidence that the protocol uses to determine the bounty rewards distribution. Both monitors and filtering ASes can also act as bounty setters, but an AS cannot act as both a monitor and a filtering AS in \sys.

\myitem{Assumptions:}
\sys assumes that ASes can authenticate the protocol messages they send (\eg setting a bounty, or claiming) in the blockchains \ie have cryptographic identities bound to their AS numbers. RIR Resource Certificates~\cite{rfc6487, rfc9323, rfc3779} already offer that. 
\sys does not require all ASes to participate in the protocol, but assumes that a set of ASes will act as monitors, \ie publish relevant portions of their routing tables. 
As a starting point, we assume that RIPE~\cite{ripe-ris} and RouteViews~\cite{routeviews} would act as aggregate monitors by exposing the views of their peers, including the routes they did not use.
Also, \sys assumes that ASes that want to get paid for filtering, \ie participate in the protocol, can evaluate BGP announcement validity \eg by having an up-to-date view of the public RPKI repositories~\cite{rfc9286}  or simply know their customers.

\subsection{Protocol Description}
\label{sec:description}
\input{figs/timeline}
We now describe the protocol. 
Because filtering is a local action that cannot be directly observed by third parties, \sys does not attempt to prove that a claimant filtered a hijack. Instead, the protocol combines filtering claims with monitor evidence to infer whether a claimant propagated a hijack, and estimate its contribution to containment. We first introduce the \textbf{messages} that \sys participants use to interact with the protocol, and then describe how they are produced throughout the protocol's execution. Reward distribution is discussed in detail in Section~\ref{sec:reward_determination}.

\myitem{bounty} messages (Fig.~\ref{fig:motivating-example}: \circled{2}) are smart contracts that encode the terms set by the bounty setter for rewarding filtering hijacks of its prefix. \bounty messages contain the protected \prefix and an associated \proofofprefixownership (\eg an RIR-issued resource certificate~\cite{rfc6487}), an AS \denylist, and escrowed \filterReward and \monitorReward balances that can only be distributed by the bounty's specific \releaseFunction (Fig.~\ref{fig:motivating-example}: \circled{8}).

\myitem{maskedClaim} messages (Fig.~\ref{fig:motivating-example}: \circled{4}) are commitments that filtering ASes publish to the blockchain to signal that they filtered an invalid announcement and intend to claim a bounty associated with it. A \maskedClaim contains the cryptographic hash of the hijack announcement's \prefix and \aspath that the filtering AS observes.

\myitem{RIBCommit} messages (Fig.~\ref{fig:motivating-example}: \circled{5}) are published by monitor ASes to commit to the state of their routing tables. A \ribCommit contains the merkle root of all (\prefix, \aspath) tuples in a monitor's routing information base (RIB).

\myitem{claim} messages (Fig.~\ref{fig:motivating-example}: \circled{6}) are published by filtering ASes to reveal the details of their previously published \maskedClaim, and claim a \bounty. \claim messages contain the filtered hijack \prefix and \aspath in clear text and a pointer to the corresponding \maskedClaim.

\myitem{subRIBCommit} (Fig.~\ref{fig:motivating-example}: \circled{7}) messages are published by monitors to reveal routing entries that are relevant to a particular \bounty. Specifically, a \subribCommit contains (\prefix, \aspath) tuples where \prefix matches that of a \claim, or \aspath contains the filtering AS from the \claim followed directly by that AS's neighbor in the \claim. The \subribCommit includes a merkle inclusion proof for each tuple that uses the merkle root from the monitor's previous \ribCommit.

\sys has five \textbf{execution phases} for each bounty that span from its publication up to reward distribution. The timeline in Figure~\ref{fig:timeline} illustrates \sys protocol phases and the messages published in them by each participant. We now describe what \sys participants do in each protocol execution phase.

\myitem{Bounty Publication:} The bounty setter who manages \prefix publishes a \bounty contract on the blockchain. Filtering ASes observe the \bounty, verify its authenticity via the \proofofprefixownership, confirm their eligibility to claim it by not being listed in the \denylist, and review the reward terms in \releaseFunction. They then install filters on their routers in anticipation of observing an invalid BGP announcement for \prefix.

\myitem{Claims Commit:} When filtering ASes filter an invalid announcement for \prefix, they create their \maskedClaim messages using the announcement's \prefix and \aspath, and immediately post them to the blockchain. The Claims Commit phase for the specific \bounty starts when the first \maskedClaim is posted and concludes after $\Delta_{\text{commit}}$.

\myitem{RIBs Commit:} This phase runs in parallel to the Claims Commit phase, and during it, route monitors observe the blockchain and each publish a \ribCommit message on the chain in response to new \maskedClaim messages submitted by filtering ASes.
To avoid over-burdening monitors, \sys requires monitors to only post a new \ribCommit if either a $\Delta_{\text{cooldown}}$ has elapsed since their last \ribCommit, or if any route containing a claimant AS has changed since the previous \ribCommit. This phase ends concurrently with the Claims Commit phase.

\myitem{Claims Reveal:} A filtering AS can reveal its claim by posting a \claim message after $\Delta_{\text{commit}}$ has elapsed since it published its \maskedClaim. The first AS to do so for this \bounty starts the Claims Reveal phase in which all other filtering ASes must submit their \claim before the phase concludes. 
This phase concludes after $\Delta_{\text{reveal}}$.

\myitem{Evidence Submission:} During this phase monitors identify the published \claim messages, and submit a \subribCommit that includes merkle proofs they construct using their routing tables and merkle roots from their previous \ribCommit. The evidence submission phase lasts for $\Delta_{\text{evidence}}$. 

\myitem{Reward Distribution}: When the Evidence Submission phase concludes, the \releaseFunction executes to compute reward allocations and distribute the \filterReward and \monitorReward to filtering ASes, and monitors respectively. We discuss the \releaseFunction claim evaluation logic in \S~\ref{sec:claim_evaluation} and the reward determination logic in \S~\ref{sec:reward_determination}.

\subsection{Properties}
\label{sec:properties}

\myitem{Trustless Reward Distribution:} \sys distributes bounty payments solely based on the bounty payment policy specified by the bounty setter, and public protocol evidence. Consequently, filtering ASes do not need to trust bounty setters to evaluate their claims fairly, and bounty setters do not need to trust the claimants to report their filtering honestly. This is enabled by encoding the bounty policy as a smart contract that automatically evaluates claims and monitor evidence to distribute the escrowed rewards.

\myitem{False Claim Resistance:} Accepted claims are bound to information committed by the claimant before they learn the contents of other claims.
Thus, ASes cannot create copycat filtering claims by viewing published claims and adapting their information.
\sys achieves this through the commit-reveal scheme described in \S~\ref{sec:description} which separates claim commitment from disclosure.

\myitem{Evidence Integrity:} Monitor evidence used to evaluate claims cannot be modified after claim content becomes public. As a result, monitors cannot retroactively fabricate routing evidence to supports fraudulent claims or contradicts honest ones.
\sys design attains this by requiring monitors to commit to their routing tables before claims are revealed, and subsequently prove consistency using Merkle inclusion proofs.

\myitem{Public Visibility:} All protocol messages and reward decisions in \sys are publicly visible. Bounty setters can thus identify misbehaving participants, such as claimants repeatedly submitting fraudulent claims or monitors that consistently fail to submit evidence, and exclude them claiming bounties. \sys does this by recording all protocol interactions on a public blockchain. 

\myitem{Incentive Compatibility:} 
\sys exposes configurable reward parameters that allow bounty setters to align AS incentives with deploying and performing route filtering. By configuring rewards that cover the operational costs of participating in the protocol and filtering invalid announcements, bounty setters can make honest filtering economically preferable to propagation or non-participation. At the same time, fraudulent claims are unlikely to receive substantial rewards, and can lead to the claimant becoming ineligible to claim bounties. Our protocol design achieves this using configurable bounty parameters, evidence-based reward distribution, and participant exclusion lists.

%% file: figs/timeline.tex
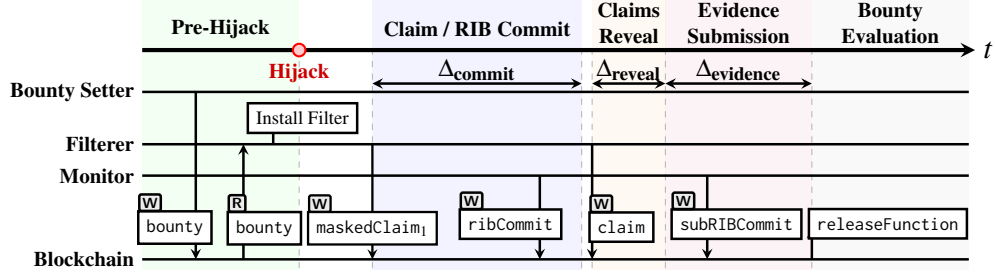
\begin{figure*}[htbp]
\centering
\resizebox{0.75\linewidth}{!}{
\begin{tikzpicture}[
    >=stealth,
    scale=1.0,
    lane/.style={draw=black, line width=0.12em},
    phasefill/.style={opacity=0.15},
    arrow/.style={->, line width=0.12em, color=black},
    axis/.style={->, line width=0.2em},
    timearrow/.style={<->, line width=0.12em, font=\small},
    rwbox/.style 2 args={
        draw=black, 
        fill=white, 
        font=\small, 
        inner sep=0.4em, 
        align=center,
        label={[font=\scriptsize\bfseries\sffamily, draw=black, fill=gray!15, inner sep=0.15em, rounded corners=0.1em, anchor=south west, xshift=0.05em, yshift=-0.1em]north west:#1}
    },
    actionbox/.style={
        draw=black, 
        fill=white, 
        font=\small, 
        inner sep=0.4em, 
        align=center,
    },
]

    \def\xStart{0em}
    \def\xHijack{\xStart+7.5em}        
    \def\xFirstClaim{\xHijack+3.5em}   
    \def\xClaimEnd{\xFirstClaim+10em}     
    \def\xRevealStart{\xClaimEnd+0.5em} 
    \def\xRevealEnd{\xClaimEnd+4.0em}    
    \def\xEvidenceEnd{\xRevealEnd+7.0em}  
    \def\xEnd{\xEvidenceEnd+7.5em}

    \def\yTime{\ySetter+2em}
    \def\ySetter{\yFilterer+2.5em}
    \def\yFilterer{\yMonitor+1.5em}
    \def\yMonitor{\yChain+4em}
    \def\yChain{0.0em}

    \fill[green!30, phasefill] (\xStart,\yTime+2.5em) rectangle (\xHijack,\yChain-0.5em);
    \fill[blue!30, phasefill]  (\xFirstClaim,\yTime+2.5em) rectangle (\xClaimEnd,\yChain-0.5em);
    \fill[orange!30, phasefill] (\xRevealStart,\yTime+2.5em) rectangle (\xRevealEnd,\yChain-0.5em);
    \fill[purple!30, phasefill] (\xRevealEnd,\yTime+2.5em) rectangle (\xEvidenceEnd,\yChain-0.5em);
    \fill[gray!30, phasefill]   (\xEvidenceEnd,\yTime+2.5em) rectangle (\xEnd,\yChain-0.5em);

    \foreach \x in {\xHijack, \xFirstClaim, \xClaimEnd, \xRevealStart, \xRevealEnd, \xEvidenceEnd} {
        \draw[dashed, gray!60] (\x, \yTime+0.8) -- (\x, \yChain-0.5em);
    }

    \draw[axis] (\xStart, \yTime) -- (\xEnd+0.3em, \yTime) node[right, font=\Large\bfseries] {$t$};

    \draw[lane] (\xStart, \ySetter) node[left, font=\bfseries] {Bounty Setter} -- (\xEnd, \ySetter);
    \draw[lane] (\xStart, \yFilterer) node[left, font=\bfseries] {Filterer} -- (\xEnd, \yFilterer);
    \draw[lane] (\xStart, \yMonitor) node[left, font=\bfseries] {Monitor} -- (\xEnd, \yMonitor);
    \draw[lane] (\xStart, \yChain) node[left, font=\bfseries] {Blockchain} -- (\xEnd, \yChain);

    \node[font=\normalsize\bfseries, align=center] at ({(\xStart+\xHijack)/2}, \yTime+1em) {Pre-Hijack};
    \node[font=\normalsize\bfseries, align=center] at ({(\xFirstClaim+\xClaimEnd)/2}, \yTime+1em) {Claim / RIB Commit};
    \node[font=\normalsize\bfseries, align=center] at ({(\xRevealStart+\xRevealEnd)/2}, \yTime+1.3em) {Claims\\ Reveal};
    \node[font=\normalsize\bfseries, align=center] at ({(\xRevealEnd+\xEvidenceEnd)/2}, \yTime+1.3em) {Evidence\\ Submission};
    \node[font=\normalsize\bfseries, align=center] at ({(\xEvidenceEnd+\xEnd)/2}, \yTime+1.3em) {Bounty\\ Evaluation};

    \draw[red, fill=red!20, line width=0.1em] (\xHijack, \yTime) node[below=0.25em, font=\normalsize\bfseries, color=red!80!black] {Hijack} circle (0.3em);

    \draw[timearrow] (\xFirstClaim, \yTime-1.6em) -- node[fill=none, inner sep=0.1em, font=\large\bfseries, above=0.0em] {$\Delta_{\text{commit}}$} (\xClaimEnd, \yTime-1.6em);
    \draw[timearrow] (\xRevealStart, \yTime-1.6em) -- node[fill=none, inner sep=0.1em, font=\large\bfseries, above=0.0em] {$\Delta_{\text{reveal}}$} (\xRevealEnd, \yTime-1.6em);
    \draw[timearrow] (\xRevealEnd, \yTime-1.6em) -- node[fill=none, inner sep=0.1em, font=\large\bfseries, above=0.0em] {$\Delta_{\text{evidence}}$} (\xEvidenceEnd, \yTime-1.6em);

    
    \draw[arrow] (0.9, \ySetter) -- (0.9, \yChain) node[pos=0.81, xshift=-1em, rwbox={Write}] {\texttt{bounty}};
    \draw[arrow] (1.7, \yChain) -- (1.7, \yFilterer) node[pos=0.27, xshift=1em, rwbox={Read}] {\texttt{bounty}};
    \draw[arrow] (2.2, \yFilterer) -- (2.2, \yFilterer+1.5em) node[pos=0.5, xshift=1.4em, yshift=0.5em, actionbox] {Install Filter};

    \draw[arrow, line width=1.1pt] (\xFirstClaim, \yFilterer) -- (\xFirstClaim, \yChain) node[pos=0.73, rwbox={Write}] {\texttt{maskedClaim}$_1$};
    \draw[arrow] (\xFirstClaim+8em, \yMonitor) -- (\xFirstClaim+8em, \yChain) node[pos=0.58, xshift=-1.4em, rwbox={Write}] {\texttt{ribCommit}};

    \draw[arrow] (\xRevealStart, \yFilterer) -- (\xRevealStart, \yChain) node[pos=0.72, xshift=1.4em, rwbox={Write}] {\texttt{claim}};

    \draw[arrow] (\xRevealEnd+2em, \yMonitor) -- (\xRevealEnd+2em, \yChain) node[pos=0.58, xshift=+1.4em, rwbox={Write}] {\texttt{subRIBCommit}};

    \draw[arrow] (\xEvidenceEnd, \yChain) -- (\xEvidenceEnd, \yChain + 1.5em) node[right, yshift=+0.2em, xshift=-0.2em, actionbox] {\texttt{releaseFunction}};

\end{tikzpicture}
}
\caption{\sys execution phases timeline. Reads and writes to the blockchain are marked with R and W respectively.}
\label{fig:timeline}
\end{figure*}

%% file: sections/reward_determination.tex
\section{Claim Evaluation}
\label{sec:claim_evaluation}

In the Reward Distribution phase, \sys invokes the \releaseFunction defined in the \bounty to distribute bounty rewards. The \releaseFunction takes \maskedClaim, \claim, \ribCommit and \subribCommit messages as input and performs four tasks: First, it validates the submitted claims and evidence. Second, it determines which claims are eligible for rewards based on the submitted evidence. Third, it estimates each claimant's contribution to hijack containment. Finally, it distributes the bounty rewards to eligible claimants and monitors. We now describe claim validation and eligibility determination steps, and discuss contribution estimation and reward distribution in \S~\ref{sec:reward_determination}.

\myitem{Claim Validation:} The \releaseFunction validates claims by verifying their structure, calculating protocol phase bounds, and discarding claims submitted outside these bounds. Specifically, valid (\maskedClaim, \claim) pairs must be submitted by the same filtering AS that is not on the \denylist. \claim messages must match the \bounty \prefix, and the filtering AS must appear in the \claim \aspath. 
The \releaseFunction calculates bounds of the Claims / RIB Commit and Claim Reveal phases using the first \claim submitted at least $\Delta_{\text{commit}}$ after its corresponding \maskedClaim. 
The function then discards (\maskedClaim, \claim) pairs where \maskedClaim was submitted outside the Claims Commit period, or the \claim message was submitted outside the Claims Reveal period.

\myitem{Evidence Validation:} Similar to Claim Validation, the \releaseFunction only considers (\ribCommit, \subribCommit) pairs submitted by the same monitor AS that is not on the \denylist. The function then checks that entry in a \subribCommit is accompanied by a valid inclusion proof that uses the merkle root in \ribCommit. Pairs where the \ribCommit was posted outside the Claims / RIBs Commit phase or the \subribCommit was posted outside the Evidence Submission phase are discarded.

\myitem{Claim Eligibility:} The \releaseFunction determines which claims are ineligible to receive rewards by identifying claimants who did not filter.
The function does this by cross-referencing \claim and \subribCommit messages and identifying \subribCommit entries where \prefix matches the hijacked prefix and \aspath contains a claimant AS. Such evidence indicates that at least one monitor observed the claimant not filtering the hijack. These claims are discarded as fraudulent and all remaining claims are marked eligible to receive the \filterReward.

\section{Reward Distribution}
\label{sec:reward_determination}

Not all filtering ASes contribute equally to hijack containment. For example, a hijacker's upstream provider that filters can prevent propagation across a vast customer cone, whereas a customer with a significantly smaller customer cone shielf fewer ASes from observing the hijack. Consequently, \sys rewards filtering ASes proportionately to their estimated contribution to containing the hijack.

\myitem{Maximum Damage Potential (MDP)} To quantify $a$'s contribution to containing hijack $h$ we define Maximum Damage Potential $\text{MDP}(a,h)$. We first explain the intuition and then discuss how we approximate it. Intuitively, $\text{MDP}(a,h)$ is the set of ASes that observe $h$ when $a$'s filtering alone could have contained the hijack, but $a$ instead chose to propagate. Specifically, we use the following counterfactual experiment: $h$ is forwarded only along the BGP-preferred AS path from the hijacker to $a$, with each upstream AS forwarding the announcement exclusively to the next AS on that path. All other propagation is suppressed. Once the announcement reaches $a$, however, $a$ propagates it according to normal BGP export policies~\cite{gao2001stable}. $MDP(a,h)$ is then calculated using the number of ASes that observed $h$. This method allows us to attribute only the propagation that would result from $a$'s decision to propagate the hijack, independent of the filtering decisions of other ASes. 

\myitem{Approximating MDP via Proxy Sets:} Because MDP is defined over a counterfactual scenario, it cannot be directly observed during protocol execution. Measuring it by propagating hijack announcements would be harmful, while estimating it online using internet-scale BGP simulations~\cite{10.1145/1064212.1064257, qiu2006cam04} would be computationally expensive and require detailed knowledge of routing policies and AS relationships that is not generally available. However, because BGP announcements are exported based on AS business relationships, the propagation of legitimate announcements as observed through monitors serves as a proxy to how a hijack would propagate. 

Leveraging this insight, \sys constructs a \proxySet $\text{P}(a,h)$ for each eligible claim to approximate $\text{MDP}(a, h)$. $\text{P}(a,h)$ for claimant AS $a$ consists of the monitors expected to observe hijack $h$ if $a$ propagated $h$ instead of filtering it. Specifically, a monitor belongs to the proxy set if its routing table includes a route where the \aspath contains segment $<a,n>$, where $n$ is $a$'s neighbor that forwarded $h$ to $a$. Intuitively, $\text{P}(a,h)$ represents monitors that receive announcements exported by $a$ that were forwarded from $n$. If $a$ were to propagate $h$, these same monitors would be expected to observe $h$ as well. We validate our hypothesis that $|\text{P}(a,h)|$ strongly correlates with $\text{MDP}(a, h)$ in Section~\ref{sec:experiments} using simulations of real world hijack incidents. Although these simulations necessarily rely on incomplete knowledge of Internet routing policies, they are sufficient to evaluate whether proxy set size is statistically correlated with MDP. Our results show a strong correlation, suggesting that \proxySet size provides a useful approximation of $a$'s Maximum Damage Potential.

\myitem{Reward Distribution:} The releaseFunction uses the available subRIBCommit messages to calculate the proxy set of each claim and then its final reward. While, \sys allows bounty setters to define custom \releaseFunction policies, we propose a simple linear payout rule in which we split the \filterReward, between a flat share, evenly divided among eligible claimants, and a proportional share, distributed according to each claimant's proxy set size. Formally, let $\text{F}_h$ denote the set of eligible claimants for hijack $h$, and $\lambda$ the fraction of \filterReward that the bounty setter selects to reserve for the flat share of the reward. $\text{R}(a,h)$, claimant $a$'s reward for filtering $h$ is calculated as:

\begin{equation}
\text{R}(a,h) = \text{\filterReward}\cdot\left( \frac{\lambda}{|\text{F}_h|} + (1-\lambda) \frac{|\text{P}(a,h)|}{\sum_{b \in \text{F}_h} |\text{P}(b,h)|} \right)
\end{equation}

%% file: sections/evaluation.tex
\section{Evaluation}
\label{sec:experiments}

\input{figs/detection_rate}

We evaluate \sys in four areas: First, we measure its ability to detect fraudulent filtering claims by impostor ASes, and quantify impostor filtering incentives. Second, we evaluate whether it rewards filtering ASes proportionately to their contribution towards hijack containment. Third, we quantify the incentives that the protocol creates for honest participants. Finally, we study how different levels of protocol adoption affect participation incentives and overall hijack containment.

\subsection{Experimental Setup}
Unless otherwise stated, all \sys evaluation experiments use Internet-scale BGP simulations to model announcement propagation under different routing and filtering scenarios. This subsection describes the common simulation methodology and datasets used across the evaluation. 

\myitem{Simulation Methodology:} Our experiments simulate BGP announcement propagation between ASes on the Internet according to the Gao-Rexford model~\cite{gao2001stable} until a stable routing state is reached. Each AS in our simulation only exports its selected best path. We break ties in the BGP best path selection algorithm uniformly at random. We use the CAIDA AS relationships dataset~\cite{caida-as-relationships} to construct the simulated topology.

\myitem{Datasets:} Our simulations use 1,018 unique hijacker ASes extracted from 7,988 real prefix hijack events identified by Cloudflare Radar~\cite{cloudflare-radar-cloudflare-radar} between December 2025 and May 2026 and assigned a high confidence score. 
Because longer-prefix hijacks are preferred over the victim's announcement throughout our routing model, their propagation depends on the hijacker but not on the victim. We therefore select one observed victim per hijacker and simulate each hijacker-victim pair once for each hijack type.
By sampling by hijacker AS rather than observed events, we avoid prolific hijackers disproportionately weighting the results.
We use 799 ASes that peer with RIPE RIS~\cite{ripe-ris} and RouteViews~\cite{routeviews} collectors as monitors. We consider \sys monitors in our experiments to not further propagate the hijacks they observe.
Our experiments that study \sys deployment alongside existing filtering ASes consider the 7,384 ASes that received a perfect RoV Score by RoVista~\cite{li-2023-rov} to enforce filtering.

\myitem{Filtering rings:} 
To construct realistic scenarios where hijacks are contained, 
we define a \emph{filtering ring} around hijacker AS $H$ as a set of ASes whose joint filtering intercepts every announcement propagation path from $H$ and thereby contains the hijack. Although filtering rings may in principle contain ASes at any distance from the hijacker, we restrict our experiments to rings composed of 1-hop and 2-hop filterers. To place filterers at the second hop, we select 1-hop neighbors of $H$ to remain non-filtering so that the hijack propagates through them. Their neighbors that can receive the announcement under valley-free routing then serve as 2-hop filterers, while the remaining 1-hop neighbors of $H$ filter directly. Filtering rings provide a controlled starting point in which the hijack is fully contained near its source, allowing us to selectively relax filtering at individual ASes or sets of ASes and measure how much propagation they would otherwise permit.

\myitem{Naming Convention:} We name filtering relationships by their sequence of hops from the hijacker, with \textit{Hij} denoting the hijacker and each subsequent term denoting the relationship of the next AS. For example, \textit{Hij-Prov} denotes a 1-hop provider of the hijacker, while \textit{Hij-Peer-Cust} denotes a 2-hop AS that is a customer of a peer to the hijacker.

\subsection{Detecting Fraudulent Claims}
\label{sec:false_claim_detection}
The first question we evaluate is whether \sys can distinguish honest filtering ASes from ASes that falsely claim to have filtered a hijack. We simulate scenarios in which impostor ASes claim bounties without filtering and evaluate whether monitor evidence causes the \releaseFunction to reject their claims. 

We sample a filtering ring for each hijacker-victim pair in our dataset and then randomly choose $k \in \{2,3,4,5\}$ 1-hop and 2-hop neighbors of the hijacker (with equal probability for each distance) to act as impostors submitting fraudulent claims. For every impostor AS $a$, we define $G(a,h)$ as the set of monitors that observed the hijack with $a$ on the AS Path. We consider an impostor claim detected when $|G(a,h)| > 0$ and thus at least one monitor publishes a \subribCommit showing $a$ propagating the hijack. Because BGP only exports the preferred route per prefix, announcements propagated by one impostor might suppress announcements propagated by another before they reach monitors. Consequently, $k$ multiple simultaneous impostors reduce the visibility of each other to monitors, resulting in a more challenging detection scenario than evaluating each impostor independently.

Figure \ref{fig:gf_fraction} shows the fraction of fraudulent claims by impostor ASes that \sys identifies. The figure presents results for both longer-prefix and same-length prefix hijacks, and across different relationships of the impostor AS to the hijacker (\eg Peer to a Provider of the Hijacker: Hij-Prov-Peer). Our results show that \sys is more effective at identifying impostor ASes for longer-prefix hijacks compared to same-length prefix hijacks. This is because longer-prefix hijacks are preferred in the BGP best path selection algorithm and thus have greater visibility compared to same-length prefix hijacks that compete with the legitimate announcement by the victim AS.

Detection rate also depends strongly on the impostor's relationship to the hijacker. In particular, impostor ASes acting as 1-hop providers (Hij-Prov) and 2-hop provider-providers (Hij-Prov-Prov) are the most likely to be caught due to their large customer cone sizes and their ability to propagate hijacked routes upstream towards Tier 1 ASes. 1-hop peers (Hij-Peer) and 2-hop provider-peers (Hij-Prov-Peer) are caught at more moderate rates due to the hijack only being propagated to their own customers. Finally, 1-hop or 2-hop relationships with a customer on the path are the most challenging, with most lying claims going unobserved. This is because customer ASes typically have very small or nonexistent customer cones, making it unlikely that a monitor will witness them propagating the hijack. 

\input{figs/payment_share_by_category}
Even though \sys detects many fraudulent claims, undetected impostors may still receive rewards. We now answer the complementary question of what rewards a successful impostor can expect to obtain. To do this, we measure the containment-proportionate reward for every undetected impostor claim ($|G(a,h)| = 0$) from our simulations. We average this reward fraction over all simulations for each relationship type between impostor and hijacker. We condition our results on undetected impostor claims as the \releaseFunction does not reward detected fraudulent claims. We note that our experiments reflect favorable scenarios for the impostors because the filtering ring contains only a small number of legitimate claimants, and thus the reward is allocated among fewer ASes resulting in a larger reward for the impostor. 

Figure~\ref{fig:payment_share_by_category} shows that only successful impostors with a provider (Hij-Prov) or provider of provider relationship to the hijacker (Hij-Prov-Prov) receive any substantial fraction of the hijack containment-proportionate part of the reward. This is a result of the \releaseFunction allocating the reward according to each claimant's \proxySet size. However, the same relationships are also the most likely to be detected. Thus, the impostor ASes with the highest incentive to submit fraudulent claims are also the most likely to be identified. 

Detecting fraudulent claims using a small number of monitors is sensitive to monitors becoming unavailable or being untrusted by the bounty setter. We quantify the robustness of impostor claim detection to monitor availability by measuring the number of monitors that \sys can use to identify impostor claims. Figure \ref{fig:gf_distribution} summarizes the distribution of $G(a,h)$ for impostor claims that \sys identifies ($G(a,h)>0$). Our results show high monitor redundancy for fraudulent claims by 1-hop providers (Hij-Prov) and 2-hop provider-providers (Hij-Prov-Prov), with the median hijack being observed by at least 300 monitors regardless of the hijack type. On the other hand, impostor claims by 1-hop peers (Hij-Peer) and 2-hop provider-peers (Hij-Prov-Peer) have higher variance and lower medians, indicating that detection is more sensitive to monitor availability. We omit analysis on redundancy for customer relationships because of their low detection rates.

\subsection{Estimating Filtering Contribution}
\label{sec:contribution_estimation}

The \releaseFunction distributes the containment-proportionate component of the \filterReward according to each claim \proxySet size ($|\text{P}(a,h)|$), which approximates a claimant's Maximum Damage Potential, $\text{MDP}(a,h)$ (\S~\ref{sec:reward_determination}). We now evaluate whether $|\text{P}(a,h)|$ is correlated with $\text{MDP}(a,h)$, and whether using $|\text{P}(a,h)|$ approximates the ideal allocation based on $\text{MDP}(a,h)$.
\input{tables/max_damage_potential}

For each hijacker-victim pair, we consider each valley-free 1-hop and 2-hop relationship type to the hijacker. For each type, we sample one AS and allow the hijack to propagate only through that AS, while all other ASes in the filtering ring filter the announcement. We calculate its $\text{MDP}(a,h)$ and $|\text{P}(a,h)|$ as described in \S~\ref{sec:contribution_estimation} for both same-length prefix and longer-prefix hijacks, then use each calculate reward allocations.

Table~\ref{tab:sf_gf_pf_correlation} reports the mean $\text{MDP}(a,h)$ and $|\text{P}(a,h)|$ per filtering AS relationship to the hijacker. Using our experiment data, we calculate the Pearson correlation coefficient between $\text{MDP}(a,h)$ and $|\text{P}(a,h)|$. We find a strong correlation for longer-prefix hijacks ($r=0.893$) and moderate correlation for same-length prefix hijacks ($r=0.565$). 

We now evaluate $|\text{P}(a,h)|$-based payment calculation. We quantify this using mean allocation overlap: the percentage of total payments that is identical between proxy-based and MDP-based rewards. The complementary percentage represents the rewards that would need to be reallocated between claimants for the two reward distributions to match exactly.  

We find a mean allocation overlap of 78.4\% for longer-prefix hijacks and 70.9\% for same-length-prefix hijacks, indicating that proxy-based allocation preserves most of the ideal containment-proportionate allocation. Figure \ref{fig:payment_divergence_linear} shows this redistribution by grouping claimants by their contribution ($\text{MDP}(a,h)$) quartile, and calculating the normalized gap between the $|\text{P}(a,h)|$ and $\text{MDP}(a,h)$-based schemes.

We find that in the case of same-length prefix hijacks, this payment allocation scheme under-allocates the containment-proportional part of the bounty reward to  the most significant contributors to containment, and over-allocates to the least significant contributors. Regardless, the absolute reward differences remain small for low-contribution claimants because they receive only a small fraction of the containment-proportionate reward. For longer-prefix hijacks the divergence instead has no consistent direction, with surpluses and shortfalls roughly cancelling within each quartile, so the average gap sits near zero even though individual claimants still deviate in both directions. We discuss an alternative $|\text{P}(a,h)|$-based payment allocation scheme that increases mean allocation overlap in Appendix~\ref{sec:power_law}.

Overall, \proxySet size offers a sufficiently accurate approximation of an AS's Maximum Damage Potential, producing containment-proportionate \filterReward allocations that largely preserve the ideal MDP-based allocation.
\input{figs/payment_divergence_linear}

\subsection{Honest Participant Incentives}
\label{sec:honest_incentives}
\input{figs/filter_reward_concentration}
The \releaseFunction allocates the containment-proportionate component of the bounty reward to a claimant's estimated contribution to hijack containment. We now evaluate how this reward is distributed among filtering ASes in practice, and whether an AS's expected earnings relate to its customer cone size, a common proxy for topological influence.

We model an early deployment of \sys in which 5\% of ASes currently enforcing ROV and 5\% of previously non-filtering ASes participate in \sys. We sample existing ROV ASes by uniformly sampling from the RoVista~\cite{li-2023-rov} list, and new filtering ASes from all remaining ASes. The union of these groups represents 5\% of the total number of ASes. We simulate hijack propagation using our hijack events dataset, assuming that all RoVista ASes continue filtering invalid announcements, while only the sampled RoVista ASes and new filtering ASes participate in \sys and submit claims. We then simulate claim submission, and reward distribution using the \releaseFunction. For each AS participating in \sys, we calculate the aggregate proportionate filtering reward it earns through its claims.

Figure ~\ref{fig:filterer_reward_concentration} show that containment-proportionate rewards are highly concentrated among a small fraction of filtering ASes. Specifically, 90\% of the total proportionately allocated reward is distributed among just 4.5\% and 5.5\% of filtering ASes for longer-prefix and same-length prefix hijacks respectively. Furthermore, we find that ASes with larger customer cone sizes earn higher rewards. We find that filtering ASes in top 1\% by customer cone size earn on average 3.2x and 1.8x more rewards than ASes in the next 9\% range for same-length and longer-prefix hijacks, respectively. Consistent with this trend, as shown in Fig.~\ref{fig:filterer_reward_concentration}, the majority of rewards are allocated to filtering ASes in the 1-10\% range. Inspection of the highest-earning ASes further supports this finding: the ten ASes receiving the largest aggregate containment-proportionate rewards are primarily established regional transit providers.

These observations follow directly from the reward function. As discussed in \S~\ref{sec:contribution_estimation}, filtering ASes with provider (Hij-Prov) or provider of provider (Hij-Prov-Prov) relationships to the hijacker contribute the most to hijack containment and consequently receive the largest proportional rewards. Because these topological positions are occupied by a smaller set of ASes in the internet hierarchy, rewards are similarly concentrated.

\subsection{Deployment Dynamics}
\input{tables/mean_damage_reduction}
As more ASes join \sys and filter invalid announcements, we expect hijacks to become increasingly contained. At the same time, protocol adoption changes the set of ASes eligible to claim bounties, and by extension the number of eligible to receive rewards. We therefore evaluate how increasing protocol adoption affects both hijack containment and the number of eligible claimants.

We simulate \sys deployments using the methodology from \S~\ref{sec:honest_incentives}, and vary the fraction of new filtering ASes that join the protocol. For each deployment level, we simulate hijack propagation, claim submission, and reward distribution, and measure both the mean hijack spread ($|D|$) and the average number of eligible claims per bounty.

Table~\ref{tab:containment_by_setB} shows how increased protocol adoption reduces the mean number of ASes that witness a hijack ($|D|)$. In early adoption stages (5-25\% new filtering ASes range), the average number of ASes that claim a bounty increases as more ASes become eligible to receive rewards. Beyond this point, the reduced hijack visibility results in fewer average bounty claimants.

%% file: figs/detection_rate.tex
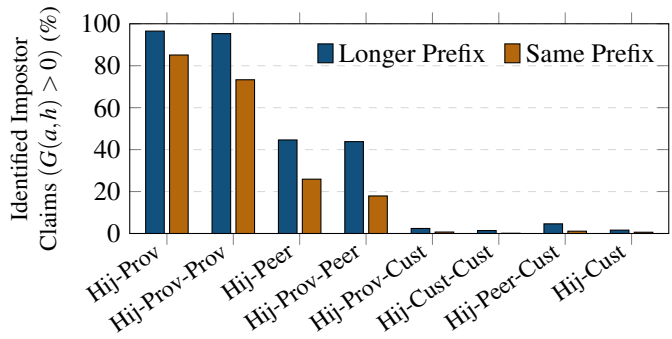
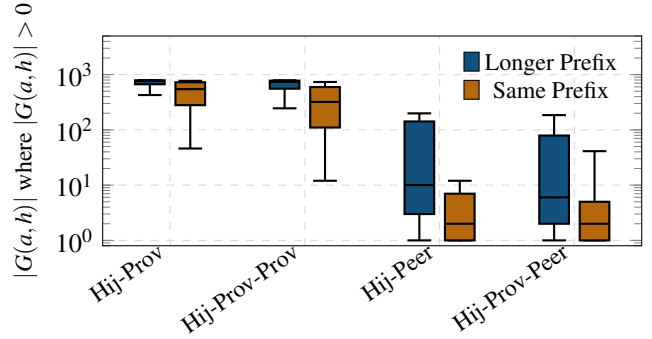
\begin{figure*}
    \centering
    \begin{subfigure}[t]{0.49\linewidth}
        \centering
            \vspace*{2.2mm} 
            \begin{tikzpicture}[baseline=(current bounding box.north)]
            \begin{axis}[
                anchor=north west,
                ybar,
                bar width=0.7em,
                width=\linewidth,
                height=0.5\linewidth,
                legend style={
                    at={(1,0.85)},
                    anchor=east,
                    legend columns=-1,
                    draw=none,
                    fill=none,
                    /tikz/every even column/.append style={column sep=0.5em}
                },
                legend image code/.code={
                    \draw[#1] (0cm,-0.1cm) rectangle (0.2cm,0.15cm);
                },
                ylabel={{Identified Impostor \\ Claims $(G(a,h)>0)$ (\%)}},
                ylabel style={font=\small, align=center},
                ymin=0,
                ymax=100,
                ymajorgrids=true,
                grid style={dashed, gray!30},
                symbolic x coords={
                    Hij-Prov,
                    Hij-Prov-Prov,
                    Hij-Peer,
                    Hij-Prov-Peer,
                    Hij-Prov-Cust,
                    Hij-Cust-Cust,
                    Hij-Peer-Cust,
                    Hij-Cust
                },
                xtick=data,
                x tick label style={rotate=35, anchor=east, font=\small},
                enlarge x limits=0.08
            ]

            \addplot+[
                fill=hexBlue,
                draw=black,
            ] table[x=Category, y=LongerPrefix, col sep=comma] {data/gf_data.csv};

            \addplot+[
                fill=hexBronze,
                draw=black,
            ] table[x=Category, y=SamePrefix, col sep=comma] {data/gf_data.csv};

            \legend{Longer Prefix, Same Prefix}

            \end{axis}
            \end{tikzpicture}
        \caption{Fraction of impostor claims that \sys identified by at least one monitor $(G(a,h) > 0)$ per type of hijacker-impostor relationship.}
        \label{fig:gf_fraction}
    \end{subfigure}
    \hfill
    \begin{subfigure}[t]{0.49\linewidth}
        \centering
        \vspace*{0.0mm}   
        \begin{tikzpicture}[baseline=(current bounding box.north)]
            \begin{axis}[
                anchor=north west,
                ylabel={$|G(a,h)|$ where $|G(a,h)| > 0$},
                ylabel style={font=\small, align=center},
                xtick={1,2,3,4},
                xticklabels={
                    Hij-Prov,
                    Hij-Prov-Prov,
                    Hij-Peer,
                    Hij-Prov-Peer
                },
                xtick style={draw=none},
                xmin=0.5, xmax=4.5,
                ymin=0.8,
                ymax=5000,
                ymajorgrids=true,
                ymode=log,
                grid style={dashed, gray!30},
                width=\linewidth,
                height=0.5\linewidth,
                grid=major,
                boxplot/draw direction=y,
                x tick label style={rotate=35, anchor=east, font=\small},
                legend style={
                    at={(0.98,0.97)},
                    anchor=north east,
                    draw=none,
                    fill=none,
                    font=\small,
                    row sep=-0.1em
                },
                legend image code/.code={
                    \draw[#1] (0cm,-0.1cm) rectangle (0.2cm,0.15cm);
                },
            ]
                \pgfplotstableread[col sep=comma]{data/gf_boxplot_data.csv}\boxplotdata
                \pgfplotstablegetrowsof{\boxplotdata}
                \pgfmathsetmacro{\boxplotlastrow}{\pgfplotsretval-1}

                \pgfplotsinvokeforeach{0,...,\boxplotlastrow}{
                    \pgfplotstablegetelem{#1}{Position}\of\boxplotdata
                    \edef\boxpos{\pgfplotsretval}
                    \pgfplotstablegetelem{#1}{Offset}\of\boxplotdata
                    \edef\boxoffset{\pgfplotsretval}
                    \pgfplotstablegetelem{#1}{Color}\of\boxplotdata
                    \edef\boxcolor{\pgfplotsretval}
                    \pgfplotstablegetelem{#1}{P10}\of\boxplotdata
                    \edef\boxpten{\pgfplotsretval}
                    \pgfplotstablegetelem{#1}{Q1}\of\boxplotdata
                    \edef\boxqone{\pgfplotsretval}
                    \pgfplotstablegetelem{#1}{Median}\of\boxplotdata
                    \edef\boxmed{\pgfplotsretval}
                    \pgfplotstablegetelem{#1}{Q3}\of\boxplotdata
                    \edef\boxqthree{\pgfplotsretval}
                    \pgfplotstablegetelem{#1}{P90}\of\boxplotdata
                    \edef\boxpninety{\pgfplotsretval}

                    \edef\boxplotopts{%
                        black, thick, fill=\boxcolor,
                        boxplot prepared={%
                            draw position={\boxpos + \boxoffset},
                            lower whisker=\boxpten,
                            lower quartile=\boxqone,
                            median=\boxmed,
                            upper quartile=\boxqthree,
                            upper whisker=\boxpninety,
                            box extend=0.22
                        },
                        forget plot%
                    }
                    \expandafter\addplot\expandafter[\boxplotopts] coordinates {};
                }

                \addlegendimage{fill=hexBlue, draw=black}
                \addlegendimage{fill=hexBronze, draw=black}
                \legend{Longer Prefix, Same Prefix}
            \end{axis}
        \end{tikzpicture}
        \vspace*{1.2mm}
        \caption{Number of monitors that identify impostor claims per type of hijacker-impostor relationship.}
        \label{fig:gf_distribution}
    \end{subfigure}

    \caption{(a) Fraudulent filtering claims for longer-prefix hijacks by providers (Hij-Prov) or providers of providers of the hijacker (Hij-Prov-Prov) are the most likely to be identified. (b) When caught, these same relationship types are also observed by hundreds of monitors rather than just one, making the claims robust even if some monitors are unavailable. For (b), whiskers extend to the 10th and 90th percentiles.}
    \label{fig:gf_evidence}
\end{figure*}

%% file: figs/payment_share_by_category.tex
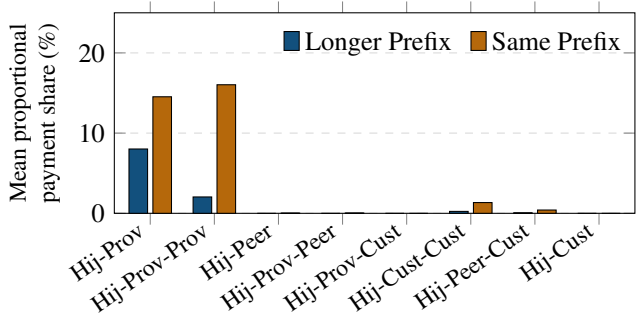
\begin{figure}[t]
    \centering
    \begin{tikzpicture}
        \begin{axis}[
            ybar,
            bar width=0.7em,
            width=\linewidth,
            height=0.5\linewidth,
            legend style={
                at={(1,0.85)},
                anchor=east,
                legend columns=-1,
                draw=none,
                fill=none,
                /tikz/every even column/.append style={column sep=0.5em}
            },
            legend image code/.code={
                \draw[#1] (0cm,-0.1cm) rectangle (0.2cm,0.15cm);
            },
            ylabel={Mean proportional\\ payment share (\%)},
            ylabel style={font=\small, align=center},
            ymin=0,
            ymax=25,
            ymajorgrids=true,
            grid style={dashed, gray!30},
            symbolic x coords={
                Hij-Prov,
                Hij-Prov-Prov,
                Hij-Peer,
                Hij-Prov-Peer,
                Hij-Prov-Cust,
                Hij-Cust-Cust,
                Hij-Peer-Cust,
                Hij-Cust
            },
            xtick=data,
            x tick label style={rotate=35, anchor=east, font=\small},
            enlarge x limits=0.08
        ]

        \addplot+[
            fill=hexBlue,
            draw=black,
        ] table[x=Category, y=LongerPrefix, col sep=comma] {data/payment_share_by_category.csv};

        \addplot+[
            fill=hexBronze,
            draw=black,
        ] table[x=Category, y=SamePrefix, col sep=comma] {data/payment_share_by_category.csv};

        \legend{Longer Prefix, Same Prefix}

        \end{axis}
    \end{tikzpicture}
    \caption{Only providers (Hij-Prov) and providers of providers (Hij-Prov-Prov) to the hijacker AS stand to gain any meaningful share of containment-proportional bounty rewards through fraudulent claims. These are also the easiest to catch by \sys.}
    \label{fig:payment_share_by_category}
\end{figure}

%% file: tables/max_damage_potential.tex
\begin{table}[t]
\centering
\small
\caption{AS proxy set size ($|\text{P}(a,h)|$) correlates strongly with Maximum Damage Potential ($|MDP(a,h)|$) for longer-prefix hijacks and moderately for same-prefix hijacks. This correlation enables \sys to pay claimants using $|P(a,h)|$ that can be observed through monitors rather $|D_F|$, which cannot. }
\label{tab:sf_gf_pf_correlation}

\setlength{\tabcolsep}{4pt}
\resizebox{\linewidth}{!}{%
\begin{tabular}{l!{\vrule width \lightrulewidth}rr!{\vrule width \lightrulewidth}rr!{\vrule width \lightrulewidth}r}
\toprule
\multicolumn{1}{c!{\vrule width \lightrulewidth}}{
    \multirow{2}{*}{\shortstack{\textbf{Relationship}\\\textbf{Type}}}
} & \multicolumn{2}{c!{\vrule width \lightrulewidth}}{\textbf{Longer Prefix}} & \multicolumn{2}{c!{\vrule width \lightrulewidth}}{\textbf{Same-length Prefix}} &  \\
\multicolumn{1}{c!{\vrule width \lightrulewidth}}{} & \multicolumn{1}{c}{\textbf{$|MDP(a,h)|$}} & \multicolumn{1}{c!{\vrule width \lightrulewidth}}{\textbf{$|G(a,h)|$}} & \multicolumn{1}{c}{\textbf{$|MDP(a,h)|$}} & \multicolumn{1}{c!{\vrule width \lightrulewidth}}{\textbf{$|G(a,h)|$}} & \multicolumn{1}{c}{\textbf{$|P(a,h)|$}} \\
\midrule
Hij-Prov & 72,256.1 & 721.1 & 13,458.5 & 468.8 & 648.1 \\
Hij-Peer & 6,749.8 & 57.5 & 89.9 & 2.0 & 5.4 \\
Hij-Cust & 119.3 & 1.0 & 2.4 & 0.0 & 0.1 \\
Hij-Prov-Prov & 66,255.6 & 643.5 & 3,319.5 & 239.9 & 454.9 \\
Hij-Prov-Peer & 5,555.2 & 47.2 & 30.3 & 1.1 & 3.2 \\
Hij-Prov-Cust & 159.1 & 1.3 & 4.2 & 0.0 & 0.0 \\
Hij-Peer-Cust& 142.4 & 1.3 & 2.6 & 0.0 & 0.0 \\
Hij-Cust-Cust & 3.6 & 0.0 & 2.0 & 0.0 & 0.0 \\
\bottomrule
\end{tabular}
}

\vspace{4pt}
\centering
\footnotesize
\textbf{Pearson $r(|MDP(a,h)|, |P(a,h)|)$:} \textbf{0.893} (longer), \textbf{0.565} (same-length)
\end{table}

%% file: figs/payment_divergence_linear.tex
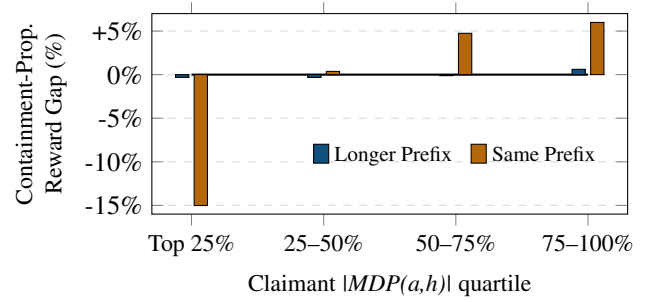
\begin{figure}[t]
    \centering
    \begin{tikzpicture}[baseline=(current bounding box.north)]
        \begin{axis}[
            anchor=north west,
            ybar,
            bar width=5pt,
            width=0.93\linewidth,
            height=0.5\linewidth,
            ylabel={Containment-Prop.\\ Reward Gap (\%)},
            ylabel style={font=\small, align=center},
            xlabel={Claimant \emph{|MDP(a,h)|} quartile},
            xlabel style={font=\small, align=center},
            ymin=-16, ymax=7,
            ytick={-15,-10,-5,0,5},
            yticklabels={-15\%,-10\%,-5\%,0\%,+5\%},
            ymajorgrids=true,
            grid style={dashed, gray!30},
            symbolic x coords={Top25,25to50,50to75,Bottom25},
            xtick=data,
            xticklabels={Top 25\%, 25\textendash 50\%, 50\textendash 75\%, 75\textendash 100\%},
            x tick label style={font=\small},
            legend style={
                at={(axis cs:Bottom25,-10)},
                anchor=north east,
                xshift=8pt,
                yshift=10pt,
                legend columns=2,
                legend cell align={left},
                draw=none,
                fill=none,
                font=\footnotesize,
                row sep=-0.1em,
                /tikz/every even column/.append style={column sep=0.5em}
            },
            legend image code/.code={
                \draw[#1] (0cm,-0.1cm) rectangle (0.2cm,0.15cm);
            },
        ]
        \draw[black, thick] (axis cs:Top25,0) -- (axis cs:Bottom25,0);

        \addplot+[fill=hexBlue, draw=black] table[x=Quartile, y=Linear, col sep=comma] {data/payment_divergence_longer.csv};
        \addplot+[fill=hexBronze, draw=black] table[x=Quartile, y=Linear, col sep=comma] {data/payment_divergence_same.csv};

        \legend{Longer Prefix, Same Prefix}
        \end{axis}
    \end{tikzpicture}

    \vspace{0.3em}

    \caption{Containment-proportionate rewards for same-length prefix hijacks calculated using $|\text{P}(a,h)|$ underpay ASes contributing more to containment, and overpay those contributing less, proving that  $|\text{P}(a,h)|$ is a good proxy of hijack containment, which is hard to calculate in practice.}
    \label{fig:payment_divergence_linear}
\end{figure}

%% file: figs/filter_reward_concentration.tex
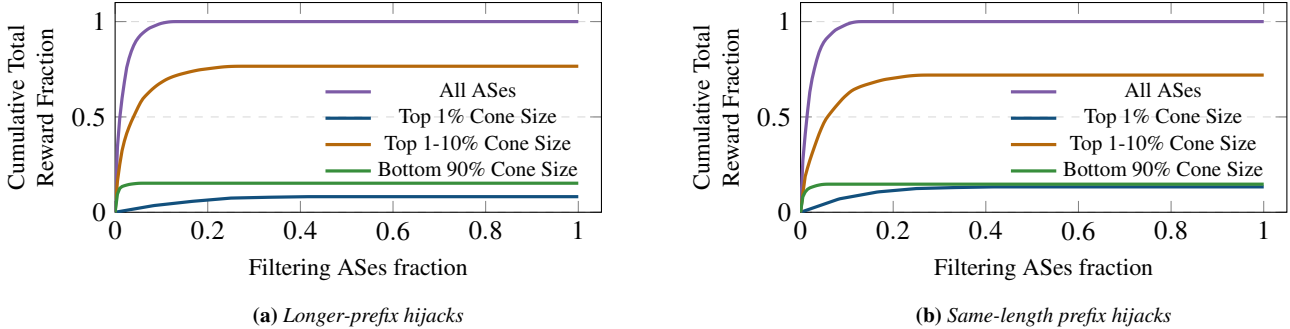
\begin{figure*}[h]
    \definecolor{hexPurple}{HTML}{7d5ba6}
    \centering
    \begin{subfigure}[t]{0.49\linewidth}
        \centering
        \begin{tikzpicture}[baseline=(current bounding box.north), trim axis left, trim axis right]
            \begin{axis}[
                anchor=north west,
                width=0.92\linewidth,
                height=0.5\linewidth,
                xlabel={Filtering ASes fraction},
                xlabel style={font=\small, align=center},
                ylabel={Cumulative Total \\ Reward Fraction},
                ylabel style={font=\small, align=center},
                xmin=0, xmax=1.05, ymin=0, ymax=1.1,
                ymajorgrids=true,
                grid style={dashed, gray!30},
                legend style={
                    at={(0.98,0.1)},
                    anchor=south east,
                    draw=none,
                    fill=none,
                    font=\footnotesize,
                    row sep=-0.1em,
                },
            ]

            \addplot+[line width=1.2pt, hexPurple, no marks] table[x=x, y=Combined, col sep=comma] {data/filterer_reward_concentration_longer.csv};
            \addplot+[line width=1.2pt, hexBlue, no marks] table[x=x, y=Tier1, col sep=comma] {data/filterer_reward_concentration_longer.csv};
            \addplot+[line width=1.2pt, hexBronze, no marks] table[x=x, y=Tier2, col sep=comma] {data/filterer_reward_concentration_longer.csv};
            \addplot+[line width=1.2pt, hexGreen, no marks] table[x=x, y=Tier3, col sep=comma] {data/filterer_reward_concentration_longer.csv};

            \legend{All ASes, Top 1\% Cone Size, Top 1-10\% Cone Size, Bottom 90\% Cone Size}
            \end{axis}
        \end{tikzpicture}
        \caption{Longer-prefix hijacks}
        \label{fig:filterer_reward_concentration_longer}
    \end{subfigure}
    \hfill
    \begin{subfigure}[t]{0.49\linewidth}
        \centering
        \begin{tikzpicture}[baseline=(current bounding box.north), trim axis left, trim axis right]
            \begin{axis}[
                anchor=north west,
                width=0.92\linewidth,
                height=0.5\linewidth,
                xlabel={Filtering ASes fraction},
                xlabel style={font=\small, align=center},
                ylabel={Cumulative Total \\ Reward Fraction},
                ylabel style={font=\small, align=center},
                xmin=0, xmax=1.05, ymin=0, ymax=1.1,
                ymajorgrids=true,
                grid style={dashed, gray!30},
                legend style={
                    at={(0.98,0.1)},
                    anchor=south east,
                    draw=none,
                    fill=none,
                    font=\footnotesize,
                    row sep=-0.1em,
                },
            ]

            \addplot+[line width=1.2pt, hexPurple, no marks] table[x=x, y=Combined, col sep=comma] {data/filterer_reward_concentration_same.csv};
            \addplot+[line width=1.2pt, hexBlue, no marks] table[x=x, y=Tier1, col sep=comma] {data/filterer_reward_concentration_same.csv};
            \addplot+[line width=1.2pt, hexBronze, no marks] table[x=x, y=Tier2, col sep=comma] {data/filterer_reward_concentration_same.csv};
            \addplot+[line width=1.2pt, hexGreen, no marks] table[x=x, y=Tier3, col sep=comma] {data/filterer_reward_concentration_same.csv};

            \legend{All ASes, Top 1\% Cone Size, Top 1-10\% Cone Size, Bottom 90\% Cone Size}
            \end{axis}
        \end{tikzpicture}
        \caption{Same-length prefix hijacks}
        \label{fig:filterer_reward_concentration_same}
    \end{subfigure}

    \caption{The majority of containment-proportionate filtering rewards are awarded to the top 1-10\% filtering ASes by cone size.}
    \label{fig:filterer_reward_concentration}
\end{figure*}

%% file: tables/mean_damage_reduction.tex
\begin{table*}[t]
\centering
\small
\caption{As more filtering ASes join the protocol hijack propagation
($D$) is significantly reduced, which also eventually causes a decrease
in the average number of claimant ASes per hijack.}
\label{tab:containment_by_setB}

\setlength{\tabcolsep}{5pt}
\begin{tabular}{l!{\vrule width \lightrulewidth}rrr!{\vrule width \lightrulewidth}rrr}
\toprule
\multicolumn{1}{c!{\vrule width \lightrulewidth}}{\multirow{2}{*}{\shortstack[l]{\textbf{New Filtering}\\\textbf{ASes (\%)}}}} & \multicolumn{3}{c!{\vrule width \lightrulewidth}}{\textbf{Longer Prefix}} & \multicolumn{3}{c}{\textbf{Same-length Prefix}} \\
\multicolumn{1}{c!{\vrule width \lightrulewidth}}{} & \multicolumn{1}{c}{\shortstack{\textbf{Mean $|D|$}\\{}}} & \multicolumn{1}{c}{\shortstack{\textbf{Reduction (\%)}\\{}}} & \multicolumn{1}{c!{\vrule width \lightrulewidth}}{\shortstack{\textbf{Avg.}\\\textbf{Claimants}}} & \multicolumn{1}{c}{\shortstack{\textbf{Mean $|D|$}\\{}}} & \multicolumn{1}{c}{\shortstack{\textbf{Reduction (\%)}\\{}}} & \multicolumn{1}{c}{\shortstack{\textbf{Avg.}\\\textbf{Claimants}}} \\
\midrule
RoVista Baseline & 8{,}527.2 & --   &  28.2 & 1{,}229.5 & --   &  9.5 \\
5\%              & 7{,}259.6 & 14.9 &  82.6 & 1{,}092.1 & 11.2 & 25.4 \\
10\%             & 6{,}198.2 & 27.3 & 134.7 &   991.5   & 19.4 & 38.9 \\
15\%             & 5{,}673.6 & 33.5 & 171.8 &   920.3   & 25.2 & 49.7 \\
20\%             & 4{,}778.3 & 44.0 & 196.6 &   799.1   & 35.0 & 59.5 \\
25\%             & 3{,}826.2 & 55.1 & 180.6 &   730.7   & 40.6 & 65.0 \\
50\%             &   946.5   & 88.9 & 112.2 &   333.0   & 72.9 & 61.6 \\
\bottomrule
\end{tabular}
\end{table*}

%% file: sections/related_work.tex
\section{Related Work}



\myitem{Traditional defenses:} Existing defenses that address BGP hijacking aim to prevent unauthorized announcements, detect hijacks, or mitigate their effects.
Early prevention systems such as S-BGP~\cite{kent2000secure}, proposed securing routing using Public Key Infrastructure (PKI) to verify prefix ownership and announcement paths, but deployment studies identified significant practical challenges~\cite{kent2000deployment}.
The Resource Public Key Infrastructure (RPKI)~\cite{lepinski2012infrastructure} provides a mechanism to associate prefixes with authorized origin ASes via Route Origin Authorizations (ROAs)~\cite{rfc9582}, enabling ASes to perform Route Origin Validation (ROV) and filter invalid announcements~\cite{rfc6811}.
BGPSec~\cite{lepinski2017bgpsec}, leverages the RPKI to provide cryptographic path validation. However, its deployment faces challenges related to computational overhead, and limited benefits under partial adoption~\cite{goldberg2014taking}.
BGP-iSec~\cite{morris2024bgp} builds on BGPSec and uses transitive signatures to prevent manipulation of announcement paths and attributes.
ASPA~\cite{ietf-sidrops-aspa-profile-29}, and subsequent extensions~\cite{furuness2025securing}, enable ASes to validate announcement paths using AS relationship declarations.
Among these mechanisms, ROV is deployed at a meaningful scale~\cite{li-2023-rov}, while ASPA is emerging as a complementary path-validation mechanism~\cite{rpki-console-aspa}.

A separate line of work focuses on detecting and mitigating hijacks after they occur. 
Works like Zheng et al.~\cite{zheng2007light}, iSPY~\cite{zhang2008ispy}, and Oscilloscope~\cite{buhler2023oscilloscope} use data-plane measurements to detect BGP hijacks.
Hybrid approaches like Argus~\cite{shi2012detecting} and Hu et al.~\cite{hu2007accurate} leverage both control- and data-plane data for real-time hijack detection.
ARTEMIS~\cite{sermpezis2018artemis} uses live BGP feeds and mitigates hijacks via prefix disaggregation, and announcement outsourcing, while HEAP~\cite{schlamp2016heap} interfaces with existing detection systems to reduce false alarms using IRR, route-collector, and TLS certificate data.
HOWLR~\cite{doumanidis2026howlr} operates at the end-user level by identifying and continuously authenticating witnesses in victim prefixes to detect BGP hijacks.
These systems complement prevention mechanisms, such as ROV, by providing detection and mitigation when hijacks propagate.

\myitem{Blockchain-based defenses:} 
Blockchain-based approaches have also been proposed to strengthen BGP security~\cite{blockchain-routing-survey}. Works like the Internet Blockchain~\cite{hari2016internet}, BGPcoin~\cite{xing2018bgpcoin} and Mastilak et al.~\cite{mastilak2020enhancing} propose blockchain-based alternatives or extensions to the RPKI for managing Internet resources and validating route origins. BRVM~\cite{liu2020novel} proposes a blockchain-based method to identify routing policy violations, while RouteChain~\cite{saad2022routechain} proposes a hierarchical blockchain architecture for detecting hijacks. 

Existing prevention and detection solutions do not directly address the incentive for ASes to deploy and exercise filtering. In contrast, \sys uses the blockchain as a coordination layer for a mechanism that incentivizes hijack filtering: bounty setters offer rewards to protect their prefixes, and independent route monitors help corroborate claims made by filtering ASes.

%% file: sections/discussion.tex
\section{Discussion \& Future Work}
\label{sec:discussion}

\myitem{Dealing with strategic collisions:} A natural concern is whether BGPay itself creates opportunities for abuse.\\
\emph{First, can an AS claim a hijack that never happened?} Because BGPay reasons partly from the absence of contradictory evidence, a claimant X could in principle fabricate an announcement that it claims to have filtered. But doing so is not an anonymous false claim: X's claim publicly and permanently names the AS that allegedly originated the hijack as well as every AS on the alleged forwarding path. This sharply constrains whom X can plausibly accuse. In particular, X must effectively frame a direct neighbor, otherwise the claim also implicates intermediate ASes that know they never forwarded the route, and that neighbor must be one whose announcement could plausibly have gone unseen elsewhere. A neighbor announcing to several other ASes would normally leave independent evidence or generate other claims, making X's account inconsistent with the public record. Thus, the easiest ASes to frame are essentially neighbors whose relevant connectivity is visible only through X; these are few, have a direct business relationship with X, and cannot be exploited repeatedly without X's sequence of otherwise-unobserved incidents becoming conspicuous.\\
\emph{Second, does paying for filtering create demand for hijacking?} A hijack by itself creates no revenue for the hijacker: it creates a bounty opportunity for the ASes that receive and filter the announcement. The relevant attack therefore requires collusion: a hijacker deliberately originates an invalid route while a partner filters it and claims the bounty. Here again, BGPay's public history provides an important deterrent. The same pair repeatedly appearing as hijacker and filterer in incidents that no independent participant corroborates is permanently visible to future bounty setters. A colluding pair may therefore succeed a bounded number of times, but once detected can be placed on the bounty's denyList, eliminating future rewards while leaving the ordinary costs of mounting hijacks. Moreover, the maximum gain from any successful incident is bounded by the bounty that the prefix owner chose to escrow.
These defenses rely today on public history and exclusion rather than explicitly modeling repeated strategic interactions. Incorporating reputation and designing collusion-resistant reward mechanisms therefore remain important directions for future work.

\myitem{From incentives to equilibrium:}
\sys shows how a prefix owner can compensate an AS for filtering, but does not yet determine how large a bounty must be to induce participation. In practice, filtering costs vary across networks and include operational risk, lost transit revenue, and the cost of participating in \sys itself. A natural next step is therefore to model these heterogeneous costs and study the resulting market equilibrium: which ASes participate at a given bounty, how prefix owners should price protection, and whether a stable bounty market emerges. This is particularly important because our results show that containment value, and consequently rewards, is highly concentrated among a relatively small number of topologically influential ASes.

Further, \sys's proxy-set size provides a strong approximation of containment impact for longer-prefix hijacks, but only moderate correlation for same-length prefix hijacks, where the current linear rule can underpay high-impact filterers and overpay lower-impact ones.
\sys intentionally makes the release function programmable; alternative estimators, nonlinear rewards, evidence thresholds, replenishable bounties, and reputation-aware payments therefore form a larger mechanism-design space.

\myitem{Incentives \& Privacy for monitors:}
\sys currently assumes that a sufficient set of ASes is willing to act as monitors. Existing RIPE RIS and RouteViews peers are a natural starting point because they already expose closely related routing information. We deliberately keep the monitor role lightweight: monitors do not determine whether a claimant filtered correctly or perform any policy evaluation; they only commit to their routing state and later reveal the subset of routes relevant to a claim. This minimizes both computational burden and unnecessary disclosure of routing information. Still, relying on altruistic participation by collectors—and on their peers being willing to expose more than their currently selected routes—may not provide sufficient coverage. A natural extension is therefore to pay monitors for providing independently verifiable routing observations, much as validators in proof-of-stake systems are compensated for independently attesting to proposed state. An open question is whether monitor rewards can be designed around their marginal contribution to observability, creating incentives to fill precisely the visibility gaps that make filtering difficult to verify today.

\myitem{From hijack mitigation to inter-domain routing:} Hijack filtering is only one instance in which one AS would benefit from an action taken by another AS but has no direct way to incentivize it. Similar mechanisms could reward ASes for actions such as dropping attack traffic during a DoS attack, propagating or suppressing particular routes, or providing desired path properties. More generally, smart contracts could provide a programmable interface through which networks express what routing behavior they are willing to pay for, while independently observable routing signals determine whether payment should be released. Characterizing which routing actions are sufficiently observable and attributable to support such markets is an open problem.

%% file: sections/conclusion.tex
\section{Conclusion}
BGP hijacks are still a problem not because there is no technical solution but because independently managed ASes have no incentives to actually filter. 
This paper describes \sys, an incentives-compatible mechanism that turns BGP hijack filtering from an uncompensated public good into a paid service. 
By combining public routing observations with a commit–reveal escrow protocol, \sys rewards ASes according to their contribution to hijack containment without requiring trust between prefix owners and filterers.
Our evaluation suggests that today's monitoring infrastructure already provides sufficient visibility where filtering matters most, making incentive-aligned BGP security a promising and practical direction.

\printendnotes

%% file: sections/appendix.tex
\section{Power-Rule Payment Allocation}
\label{sec:power_law}

Section \ref{sec:contribution_estimation} considers a \textbf{linear} payment rule for the containment-proportionate reward, in which each claimant is paid proportionally to $|P(a,h)|$. This approximates the ideal MDP-based allocation, which pays claimants proportionally to their true $|MDP(a,h)|$. A bounty owner may instead prefer a more even allocation by defining the target payment proportional to $|MDP(a,h)|^\alpha$ for $\alpha<1$, which compresses differences between high- and low-impact claimants. In this case, $|P_F|^1$ no longer approximates the compressed oracle well.

We address this by evaluating a payment scheme based on the \textbf{power} payment rule that lets bounty owners reward more even distributions of the containment-proportionate reward, replacing $|P(a,h)|$ with $|P(a,h)|^\beta$. Conceptually, this payment scheme targets a policy that distributes reward proportionally to $\mathrm{MDP}(a,h)^\alpha$. At $\alpha=1$, reward would be split strictly proportionally to true Maximum Damage Potential, so claimants with the largest customer cones capture most of it, while lower values of $\alpha$ compress this distribution, spreading reward across more claimants rather than concentrating it on whichever claimant prevented the most damage. 

For the \textbf{power} payment rule, we fit an exponent $\beta_\alpha$ on $P_F$. We consider $\alpha \in {1, \tfrac{1}{2}, \tfrac{1}{4}}$ and fit $\beta_\alpha$ by minimizing the distance between the proxy-based shares and the corresponding MDP-based target shares over a 70\% calibration subset $\mathcal{E}_{\mathrm{cal}}$ of sampled hijack events. We fit $\beta_\alpha$ by minimizing

\begin{equation}
\beta_\alpha
=
\arg\min_{\beta>0}
\sum_{e\in\mathcal{E}_{\mathrm{cal}}}
\sum_{F\in\mathcal{F}_e}
\left|
\frac{|D_{e,F}|^\alpha}
{\sum_{J\in\mathcal{F}_e}|D_{e,J}|^\alpha}
-
\frac{|P_{e,F}|^\beta}
{\sum_{J\in\mathcal{F}_e}|P_{e,J}|^\beta}
\right|,
\end{equation}

We restrict the objective to claimants with $P_F > 0$ since a proxy value of zero can't be fit to any exponent and provides no information for fitting. Thus, these zero-proxy claimants still receive the flat-rate payment but no additional proxy-weighted share. As shown in Fig. \ref{fig:zero_proxy_cdf}, 95\% of zero-proxy claimants have a counterfactual attack reach ∣MDP∣ below 99.0 for longer-prefix hijacks and below 9.0 for same-prefix hijacks, indicating that they have little impact and would receive only a small share under the MDP-based allocation anyway. For each hijack, we sample up to 10 1-hop and 2-hop claimants with $P_F > 0$ and fit $\beta$ on $\mathcal{E}_{\mathrm{cal}}$. The resulting fit yields exponents of 0.5715, 0.3705, and 0.2089 for longer-prefix hijacks at $\alpha=1$, $\tfrac12$, and $\tfrac14$ respectively, and 0.9522, 0.4858, and 0.2507 for same-prefix hijacks.

In Fig. \ref{fig:payment_divergence}, we evaluate how closely proxy-based payments match the corresponding MDP-based target allocation on the held-out set. The linear rule achieves the lowest mean allocation overlap for longer-prefix hijacks and ties for the lowest for same-prefix hijacks. As discussed in Section \ref{sec:contribution_estimation}, mean quartile-level allocation gaps near zero can mask larger claimant-level errors that cancel within a quartile and therefore do not imply closer overall agreement. In contrast, the power rule achieves progressively higher mean allocation overlap as $\alpha$ decreases, since compressing the MDP-based target reduces disparities between high- and low-impact claimants. Across all power-rule settings, the remaining error is directional, with the highest-impact quartile being underpaid relative to the MDP-based target, while the lowest-impact quartile being overpaid.
\input{figs/zero_proxy_cdf}

\input{figs/payment_divergence}

Notably, we believe this compression is not purely a weakness of the proxy-based scheme. Since low-impact filterers are consistently overpaid relative to their true contribution, the proxy-based scheme may provide additional incentive for ASes that would otherwise contribute little to still participate in filtering.

%% file: figs/zero_proxy_cdf.tex
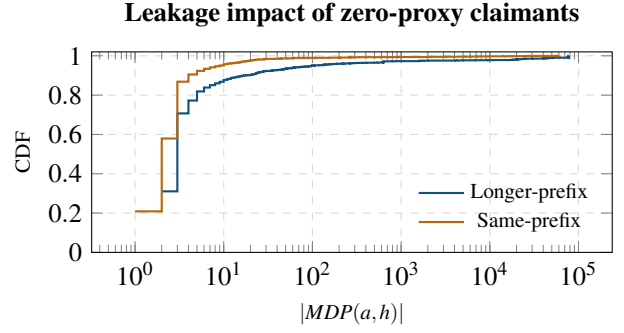
\begin{figure}[t]
    \centering
    \begin{tikzpicture}[trim axis left, trim axis right]
        \begin{axis}[
            title={\bfseries Leakage impact of zero-proxy claimants},
            title style={font=\normalsize},
            width=\linewidth,
            height=0.5\linewidth,
            xmode=log,
            xlabel={$|MDP(a,h)|$},
            ylabel={CDF},
            ylabel style={font=\footnotesize},
            xlabel style={font=\footnotesize},
            ymin=0, ymax=1.02,
            ymajorgrids=true,
            xmajorgrids=true,
            grid style={dashed, gray!30},
            legend style={
                at={(0.98,0.05)},
                anchor=south east,
                draw=none,
                fill=none,
                font=\footnotesize,
            },
        ]

        \addplot+[const plot, thick, hexBlue, no marks]
            table[x=SF, y=CDF, col sep=comma] {data/zero_proxy_cdf_longer.csv};
        \addplot+[const plot, thick, hexBronze, no marks]
            table[x=SF, y=CDF, col sep=comma] {data/zero_proxy_cdf_same.csv};

        \legend{Longer-prefix, Same-prefix}
        \end{axis}
    \end{tikzpicture}
    \caption{Both hijack types show that the vast majority of zero-proxy claimants have low Maximum Damage Potential (MDP), so excluding them from proxy-weighted payments discards little meaningful impact.}
    \label{fig:zero_proxy_cdf}
\end{figure}

%% file: figs/payment_divergence.tex
\begin{figure*}
    \centering
    \begin{subfigure}[t]{0.49\linewidth}
        \centering
        \begin{tikzpicture}[baseline=(current bounding box.north), trim axis left, trim axis right]
            \begin{axis}[
                anchor=north west,
                ybar,
                bar width=5pt,
                width=0.82\linewidth,
                height=0.55\linewidth,
                ylabel={Containment-Prop.\\Reward Gap (\%)},
                ylabel style={font=\footnotesize, align=center},
                xlabel={Claimant \emph{|MDP(a,h)|} quartile},
                xlabel style={font=\footnotesize},
                ymin=-18, ymax=9,
                ytick={-15,-10,-5,0,5},
                yticklabels={-15\%,-10\%,-5\%,0\%,+5\%},
                ymajorgrids=true,
                grid style={dashed, gray!30},
                symbolic x coords={Top25,25to50,50to75,Bottom25},
                xtick=data,
                xticklabels={Top 25\%, 25\textendash 50\%, 50\textendash 75\%, Bottom 25\%},
                x tick label style={font=\small},
                enlarge x limits=0.15,
                legend style={
                    at={(axis cs:Bottom25,-10)},
                    anchor=north east,
                    xshift=8pt,
                    yshift=10pt,
                    legend columns=2,
                    legend cell align={left},
                    draw=none,
                    fill=none,
                    font=\footnotesize,
                    row sep=-0.1em,
                    /tikz/every even column/.append style={column sep=0.5em}
                },
                legend image code/.code={
                    \draw[#1] (0cm,-0.1cm) rectangle (0.2cm,0.15cm);
                },
            ]
            \draw[black, thick] (axis cs:Top25,0) -- (axis cs:Bottom25,0);

            \addplot+[fill=gray!40, draw=black] table[x=Quartile, y=Linear, col sep=comma] {data/payment_divergence_longer.csv};
            \addplot+[fill=hexBlue, draw=blue!90!black] table[x=Quartile, y=Alpha1, col sep=comma] {data/payment_divergence_longer.csv};
            \addplot+[fill=hexBlue!65!white, draw=blue!90!black] table[x=Quartile, y=AlphaHalf, col sep=comma] {data/payment_divergence_longer.csv};
            \addplot+[fill=hexBlue!35!white, draw=blue!90!black] table[x=Quartile, y=AlphaQuarter, col sep=comma] {data/payment_divergence_longer.csv};

            \legend{Linear ($\beta=1$), $\alpha=1$, $\alpha=\tfrac{1}{2}$, $\alpha=\tfrac{1}{4}$}
            \end{axis}
        \end{tikzpicture}

        \vspace{0.3em}
        \begin{tabular}{@{}c@{}}
        {\footnotesize Mean allocation overlap:} \\
        {\footnotesize Linear: 78.4\%\quad $\alpha{=}1$: 80.2\%\quad $\alpha{=}\tfrac{1}{2}$: 83.4\%\quad $\alpha{=}\tfrac{1}{4}$: 89.1\%}
        \end{tabular}

        \caption{Longer-prefix hijacks}
        \label{fig:payment_divergence_longer}
    \end{subfigure}
    \hfill
    \begin{subfigure}[t]{0.49\linewidth}
        \centering
        \begin{tikzpicture}[baseline=(current bounding box.north), trim axis left, trim axis right]
            \begin{axis}[
                anchor=north west,
                ybar,
                bar width=5pt,
                width=0.82\linewidth,
                height=0.55\linewidth,
                ylabel={Containment-Prop.\\Reward Gap (\%)},
                ylabel style={font=\footnotesize, align=center},
                xlabel={Claimant \emph{|MDP(a,h)|} quartile},
                xlabel style={font=\footnotesize},
                ymin=-18, ymax=9,
                ytick={-15,-10,-5,0,5},
                yticklabels={-15\%,-10\%,-5\%,0\%,+5\%},
                ymajorgrids=true,
                grid style={dashed, gray!30},
                symbolic x coords={Top25,25to50,50to75,Bottom25},
                xtick=data,
                xticklabels={Top 25\%, 25\textendash 50\%, 50\textendash 75\%, Bottom 25\%},
                x tick label style={font=\small},
                enlarge x limits=0.15,
                legend style={
                    at={(axis cs:Bottom25,-10)},
                    anchor=north east,
                    xshift=8pt,
                    yshift=10pt,
                    legend columns=2,
                    legend cell align={left},
                    draw=none,
                    fill=none,
                    font=\footnotesize,
                    row sep=-0.1em,
                    /tikz/every even column/.append style={column sep=0.5em}
                },
                legend image code/.code={
                    \draw[#1] (0cm,-0.1cm) rectangle (0.2cm,0.15cm);
                },
            ]
            \draw[black, thick] (axis cs:Top25,0) -- (axis cs:Bottom25,0);

            \addplot+[fill=gray!40, draw=black] table[x=Quartile, y=Linear, col sep=comma] {data/payment_divergence_same.csv};
            \addplot+[fill=hexBronze, draw=orange!95!black] table[x=Quartile, y=Alpha1, col sep=comma] {data/payment_divergence_same.csv};
            \addplot+[fill=hexBronze!65!white, draw=orange!95!black] table[x=Quartile, y=AlphaHalf, col sep=comma] {data/payment_divergence_same.csv};
            \addplot+[fill=hexBronze!35!white, draw=orange!95!black] table[x=Quartile, y=AlphaQuarter, col sep=comma] {data/payment_divergence_same.csv};

            \legend{Linear ($\beta=1$), $\alpha=1$, $\alpha=\tfrac{1}{2}$, $\alpha=\tfrac{1}{4}$}
            \end{axis}
        \end{tikzpicture}

        \vspace{0.3em}
        \begin{tabular}{@{}c@{}}
        {\footnotesize Mean allocation overlap:} \\
        {\footnotesize Linear: 70.9\%\quad $\alpha{=}1$: 70.9\%\quad $\alpha{=}\tfrac{1}{2}$: 78.6\%\quad $\alpha{=}\tfrac{1}{4}$: 87.0\%}
        \end{tabular}

        \caption{Same-length prefix hijacks}
        \label{fig:payment_divergence_same}
    \end{subfigure}

    \caption{Across longer-prefix and same-length prefix hijacks, the power-rule payment allocation achieves higher mean allocation overlap with increasingly compressed MDP-based target allocations as $\alpha$ decreases.}
    \label{fig:payment_divergence}
\end{figure*}